\documentclass[a4paper,12pt,headings=big,DIV=12]{scrartcl}
\pdfoutput=1
\usepackage{amsmath,amssymb}
\usepackage{color}
\usepackage[dvipsnames, table]{xcolor}
\definecolor{blue}{rgb}{0,0,0.5}
\usepackage[colorlinks,linkcolor=blue,urlcolor=blue,citecolor=blue]{hyperref}
\usepackage{graphicx}
\addtokomafont{disposition}{\rmfamily\boldmath}
\usepackage{cite}
\usepackage{xspace}
\usepackage{relsize}
\usepackage{bm}
\usepackage[utf8]{inputenc}
\usepackage{tabularx,colortbl}
\usepackage{multirow}
\usepackage{booktabs}
\usepackage{bbold}
\usepackage{afterpage}

\allowdisplaybreaks

\DeclareOldFontCommand{\bf}{\normalfont\bfseries}{\mathbf}

\begin{document}

\begin{center}
{\Large\bfseries \boldmath
\vspace*{1.5cm}
New Physics in Rare $B$ Decays after Moriond 2021
}
\\[0.8 cm]
{\large{
Wolfgang Altmannshofer$^a$, Peter Stangl$^b$
}
\\[0.5 cm]
\small
$^a$ \textit{Department of Physics and Santa Cruz Institute for Particle Physics, \\
University of California, Santa Cruz,
1156 High Street, Santa Cruz, CA 95064,
United States} \\[0.25cm]
$^b$ \textit{Albert Einstein Center for Fundamental Physics,
Institute for Theoretical Physics,\\
University of Bern,
Sidlerstrasse 5, CH-3012 Bern,
Switzerland}
}
\\[0.5 cm]
\footnotesize
E-Mail: \texttt{%
\href{mailto:waltmann@ucsc.edu}{waltmann@ucsc.edu}, \href{mailto:stangl@itp.unibe.ch}{stangl@itp.unibe.ch}
}

\end{center}

\bigskip

\begin{abstract}\noindent
The anomalies in rare $B$ decays endure. We present results of an updated global analysis that takes into account the latest experimental input -- in particular the recent results on $R_K$ and BR$(B_s \to \mu^+\mu^-)$ -- and that qualitatively improves the treatment of theory uncertainties. Fit results are presented for the Wilson coefficients of four-fermion contact interactions. We find that muon specific Wilson coefficients $C_9 \simeq -0.73$ or $C_9 = -C_{10} \simeq -0.39$ continue to give an excellent description of the data. If only theoretically clean observables are considered, muon specific $C_{10} \simeq 0.60$ or $C_9=-C_{10} \simeq -0.35$ improve over the Standard Model by $\sqrt{\Delta \chi^2} \simeq 4.7\sigma$ and $\sqrt{\Delta \chi^2} \simeq 4.6\sigma$, respectively. In various new physics scenarios we provide predictions for lepton flavor universality observables and CP asymmetries that can be tested with more data. We update our previous combination of ATLAS, CMS, and LHCb data on BR$(B_s \to \mu^+\mu^-)$ and BR$(B^0\to \mu^+\mu^-)$  taking into account the full two-dimensional non-Gaussian experimental likelihoods.
\end{abstract}

\section{Introduction}

Since several years there exist persistent discrepancies between the Standard Model (SM) predictions and the experimental results for rare decays of $B$ mesons based on the neutral current $b \to s \ell \ell$ transitions. Discrepancies are seen in the branching fractions of the rare decays $B \to K \mu^+\mu^-$, $B \to K^* \mu^+\mu^-$, $B_s \to \phi \mu^+\mu^-$, and $B_s \to \mu^+\mu^-$, in the angular distribution of $B \to K^* \mu^+\mu^-$ and in lepton flavor universality (LFU) ratios. Of particular interest are the hints for LFU violation that have been observed by LHCb in the following ratios of branching fractions
\begin{equation}
    R_{K} = \frac{\text{BR}(B \to K \mu^+\mu^-)}{\text{BR}(B \to K e^+e^-)} ~,\qquad R_{K^{*}} = \frac{\text{BR}(B \to K^{*} \mu^+\mu^-)}{\text{BR}(B \to K^{*} e^+e^-)} ~.
\end{equation}
While the SM predictions for most absolute branching fractions and also the angular observables are potentially subject to large hadronic uncertainties, the LFU ratios $R_K$ and $R_{K^*}$ can be predicted with high accuracy. Significant deviations in these observables would thus constitute clean indirect evidence for new physics. Also the absolute branching ratio of the purely leptonic decay $B_s \to \mu^+\mu^-$ can be considered as theoretically clean. Non-perturbative physics enters through a single hadronic parameter, the $B_s$ meson decay constant, which is know with high precision from lattice QCD calculations.

Intriguingly, the simplest new physics scenarios that address the theoretically clean hints for LFU violation simultaneously explain also the other discrepancies. Parameterizing the new physics in terms of four fermion contact interactions, global fits of rare $B$ decay data find consistently very strong preference for new physics in the form of the operator $\frac{1}{\Lambda_\text{NP}^2}(\bar s \gamma_\alpha P_L b)(\bar\mu \gamma^\alpha \mu)$ or $\frac{1}{\Lambda_\text{NP}^2}(\bar s \gamma_\alpha P_L b)(\bar\mu \gamma^\alpha P_L \mu)$ with a generic new physics scale of $\Lambda_\text{NP} \simeq 35$~TeV (for recent work see~\cite{Geng:2017svp, Alguero:2019ptt, Alok:2019ufo, Ciuchini:2019usw, Datta:2019zca, Aebischer:2019mlg, Kowalska:2019ley, Hurth:2020rzx, Ciuchini:2020gvn, Hurth:2020ehu}).

Very recently, the LHCb collaboration presented updated results for two theoretically clean observables that have previously shown tensions with the SM predictions: the LFU ratio $R_K$ and the branching ratio BR$(B_s \to \mu^+\mu^-)$. Using the full run 2 data set the value for $R_K$ is~\cite{Aaij:2021vac}
\begin{equation}\label{eq:RK_new}
 R_K = 0.846{\phantom{.}}^{+0.042}_{-0.039} {\phantom{.}}^{+0.013}_{-0.012} \,, \qquad \text{for}~ 1.1\,\text{GeV}^2 < q^2 < 6\,\text{GeV}^2 \,,
\end{equation}
where the first uncertainty is statistical and the second one systematic, and $q^2$ is the di-muon invariant mass squared.
The new result has exactly the same central value as the previous result $R_K = 0.846^{+0.060}_{-0.054}{}^{+0.016}_{-0.014} $~\cite{Aaij:2019wad},
while the statistical uncertainty has been reduced by approximately $30\%$, commensurate with the increased statistics.
Consequently, the tension between the experimental measurement and the SM prediction, which is unity to an excellent approximation, has increased from previously $2.5\sigma$ to now $3.1\sigma$.

The branching ratio of the $B_s \to \mu^+\mu^-$ decay measured with the full run 2 data is found to be~\cite{LHCb:2021awg,LHCb:2021vsc}
\begin{equation}
 \overline{\text{BR}}(B_s \to \mu^+\mu^-) = \left(3.09{\phantom{.}}^{+0.46}_{-0.43} {\phantom{.}}^{+0.15}_{-0.11}\right) \times 10^{-9}~,
\end{equation}
where the first uncertainty is statistical and the second one systematic. This result by itself has a precision close to the previous world average BR$(B_s \to \mu^+\mu^-) = (2.69^{+0.37}_{-0.35})\times 10^{-9}$~\cite{LHCb:2020zud} that was based on results from ATLAS, CMS, and LHCb~\cite{Aaij:2017vad, Aaboud:2018mst, Sirunyan:2019xdu}.
Compared to the previous measurement by LHCb, BR$(B_s \to \mu^+\mu^-) = (3.0\pm0.6^{+0.3}_{-0.2})\times 10^{-9}$~\cite{Aaij:2017vad}, the new update finds nearly the same central value.
While the LHCb result is compatible with the SM prediction within $1\sigma$, the previous world average was below the SM prediction by more than $2\sigma$.
Here, we provide an update of the world average of the $B_s \to \mu^+\mu^-$ branching ratio and the correlated $B^0 \to \mu^+\mu^-$ branching ratio, taking into account the new LHCb results.
A Gaussian approximation to our combined two-dimensional likelihood is given by
\begin{align}
  \overline{\text{BR}}(B_s\to\mu^+\mu^-)_\text{exp}
  &= (2.93\pm0.35) \times 10^{-9},
 \\
 {\text{BR}}(B^0\to\mu^+\mu^-)_\text{exp}
 &= (0.56\pm0.70) \times 10^{-10},
\end{align}
with an error correlation coefficient $\rho = -0.27$. We find a one-dimensional pull with the SM predictions of $2.3\sigma$. Details on how the combination and the discrepancy with the SM are obtained are given in the appendix~\ref{app:Bsmumu}.

The main goal of this paper is to interpret the impact of the new experimental results in a model independent way, using the well established effective Hamiltonian approach. We parameterize new physics contributions by Wilson coefficients of dimension 6 interactions evaluated at the renormalization scale $\mu = 4.8$~GeV
\begin{equation}
  \mathcal H_\text{eff}
  =   \mathcal H_\text{eff}^\text{SM} - \frac{4G_F}{\sqrt{2}} V_{tb}V_{ts}^* \frac{e^2}{16\pi^2}
  \sum_{\ell=e,\mu}
  \sum_{i=9,10,S,P}
  \left(C^{bs\ell\ell}_i O^{bs\ell\ell}_i + C^{\prime bs\ell\ell}_i O^{\prime bs\ell\ell}_i\right)
  +
   \text{h.c.}\,.
\end{equation}
We consider the following set of semi-leptonic operators
\begin{align}
O_9^{bs\ell\ell} &=
(\bar{s} \gamma_{\mu} P_{L} b)(\bar{\ell} \gamma^\mu \ell)\,,
&
O_9^{\prime bs\ell\ell} &=
(\bar{s} \gamma_{\mu} P_{R} b)(\bar{\ell} \gamma^\mu \ell)\,,\label{eq:O9}
\\
O_{10}^{bs\ell\ell} &=
(\bar{s} \gamma_{\mu} P_{L} b)( \bar{\ell} \gamma^\mu \gamma_5 \ell)\,,
&
O_{10}^{\prime bs\ell\ell} &=
(\bar{s} \gamma_{\mu} P_{R} b)( \bar{\ell} \gamma^\mu \gamma_5 \ell)\,,\label{eq:O10}
\\
O_{S}^{bs\ell\ell} &= m_b
(\bar{s} P_{R} b)( \bar{\ell}  \ell)\,,
&
O_{S}^{\prime bs\ell\ell} &= m_b
(\bar{s}  P_{L} b)( \bar{\ell}  \ell)\,,\label{eq:OS}
\\
O_{P}^{bs\ell\ell} &= m_b
(\bar{s} P_{R} b)( \bar{\ell} \gamma_5 \ell)\,,
&
O_{P}^{\prime bs\ell\ell} &= m_b
(\bar{s}  P_{L} b)( \bar{\ell} \gamma_5 \ell)\,.\label{eq:OP}
\end{align}
We do not consider semi-leptonic tensor operators, because they are not generated at dimension 6 in the Standard Model Effective Field Theory (SMEFT). Similarly, in the case of the scalar operators, we will impose the following relations among the corresponding Wilson coefficients $C_S^{bs\ell\ell} = - C_P^{bs\ell\ell}$ and $C_S^{\prime \,bs\ell\ell} = C_P^{\prime\, bs\ell\ell}$, as they hold at dimension 6 in the SMEFT~\cite{Alonso:2014csa}.
We also do not consider semi-tauonic operators or 4-quark operators, as they affect the observables we consider only at the loop level~\cite{Jager:2017gal, Crivellin:2018yvo}.

A critical aspect of global fits is the treatment of theory uncertainties. In our previous studies~\cite{Altmannshofer:2014rta, Altmannshofer:2017fio, Altmannshofer:2017yso, Aebischer:2019mlg} we have evaluated theory uncertainties and their correlations for the Wilson coefficients fixed to their SM values. This is typically a good approximation as long as the best fit results are in the vicinity of the SM point. Possible exceptions are observables that have negligible uncertainties in the SM but not in the presence of new physics contributions. A prominent example of such observables are the LFU ratios $R_K$ and $R_{K^*}$. While the experimental uncertainties still dominate for $R_K$ and $R_{K^*}$, the precision of the new $R_K$ result in eq.~(\ref{eq:RK_new}) is strong motivation to improve our treatment of theory uncertainties. In this paper we incorporate the new physics dependence of the theory uncertainties for the first time in our fit.

The paper is organized as follows: in section~\ref{sec:uncertainties} we discuss in detail the improved treatment of theory uncertainties and illustrate the size of the effect in the case of LFU observables and CP asymmetries in the presence of new physics. In section~\ref{sec:fits} we collect the results of the updated global fit. We consider scenarios with one real Wilson coefficient at a time, scenarios with two real Wilson coefficients as well as scenarios with complex Wilson coefficients. Section~\ref{sec:predictions} contains new physics predictions for a number of LFU observables and CP asymmetries that can be tested with future data. We conclude in section~\ref{sec:conclusions}. Our combination of the experimental results on the $B_s \to \mu^+\mu^-$ branching ratio is described in appendix~\ref{app:Bsmumu}.

\section{Improved Treatment of Theory Uncertainties} \label{sec:uncertainties}

Our global fits are based on a $\chi^2$ function that depends on the Wilson coefficients in the effective Hamiltonian
and that takes into account both the theoretical and experimental uncertainties in terms of covariance
matrices, $\Sigma_\text{exp}$ and $\Sigma_\text{th}$
\begin{equation} \label{eq:chi2}
 \chi^2(C_i) = \Big( \vec O_\text{exp} - \vec O_\text{th}(C_i) \Big)^\text{T} \Big( \Sigma_\text{exp} +
 \Sigma_\text{th} \Big)^{-1} \Big( \vec O_\text{exp} - \vec O_\text{th}(C_i) \Big) ~.
\end{equation}
In the above expression, the $\vec O_\text{exp}$ are the measured central values of the observables of interest and $\vec O_\text{th}$ are the corresponding theory predictions that have dependence on the considered set of Wilson coefficients $C_i$. The covariance matrix $\Sigma_\text{th}$ includes uncertainties from parametric input, in particular CKM matrix elements and form factor parameters, as well as from non-factorisable power corrections. Our treatment of the non-factorisable power corrections follows~\cite{Altmannshofer:2014rta} and is summarized in appendix~\ref{app:power}.
In previous global fits, we made the assumption that the theoretical uncertainties are well described by the covariance matrix $\Sigma_\text{th}$ determined with the SM values for the Wilson coefficients and neglected possible dependence of $\Sigma_\text{th}$ on the new physics. This has the advantage that the time consuming evaluation of $\Sigma_\text{th}$ has to be performed only once.

We have developed a computationally efficient method to determine the new physics dependence of $\Sigma_\text{th}$.
The procedure is summarized in the following and described in detail in appendix~\ref{app:uncertainties}.

As the rare $B$ decay amplitudes are linear functions of the Wilson coefficients it is possible to express the branching ratios as second order polynomials in the Wilson coefficients. The coefficients of the polynomials are independent of new physics and their correlated uncertainties can be described by a covariance matrix that needs to be determined only once. The covariance matrix of the branching ratios can then be expressed in a straight forward way in terms of the covariance matrix of the polynomial coefficients and the Wilson coefficients.

The CP averaged angular observables $S_i$, the CP asymmetries $A_i$, and the LFU ratios can be written in terms of ratios of second order polynomials, while the $P_i^\prime$ observables involve also irrational functions. In those cases we obtain an approximation of the covariance matrix for the observables by expanding the functions to second order in the Wilson coefficients and then following the same procedure as for the branching ratios. We find that this procedure gives reliable estimates as long as the absolute values of the new physics Wilson coefficients are somewhat smaller than the corresponding relevant SM coefficients. In principle, the accuracy of the approximation could be systematically improved by expanding to higher orders.

\bigskip

\begin{figure}[tb]
\centering
\includegraphics[width=0.49\textwidth]{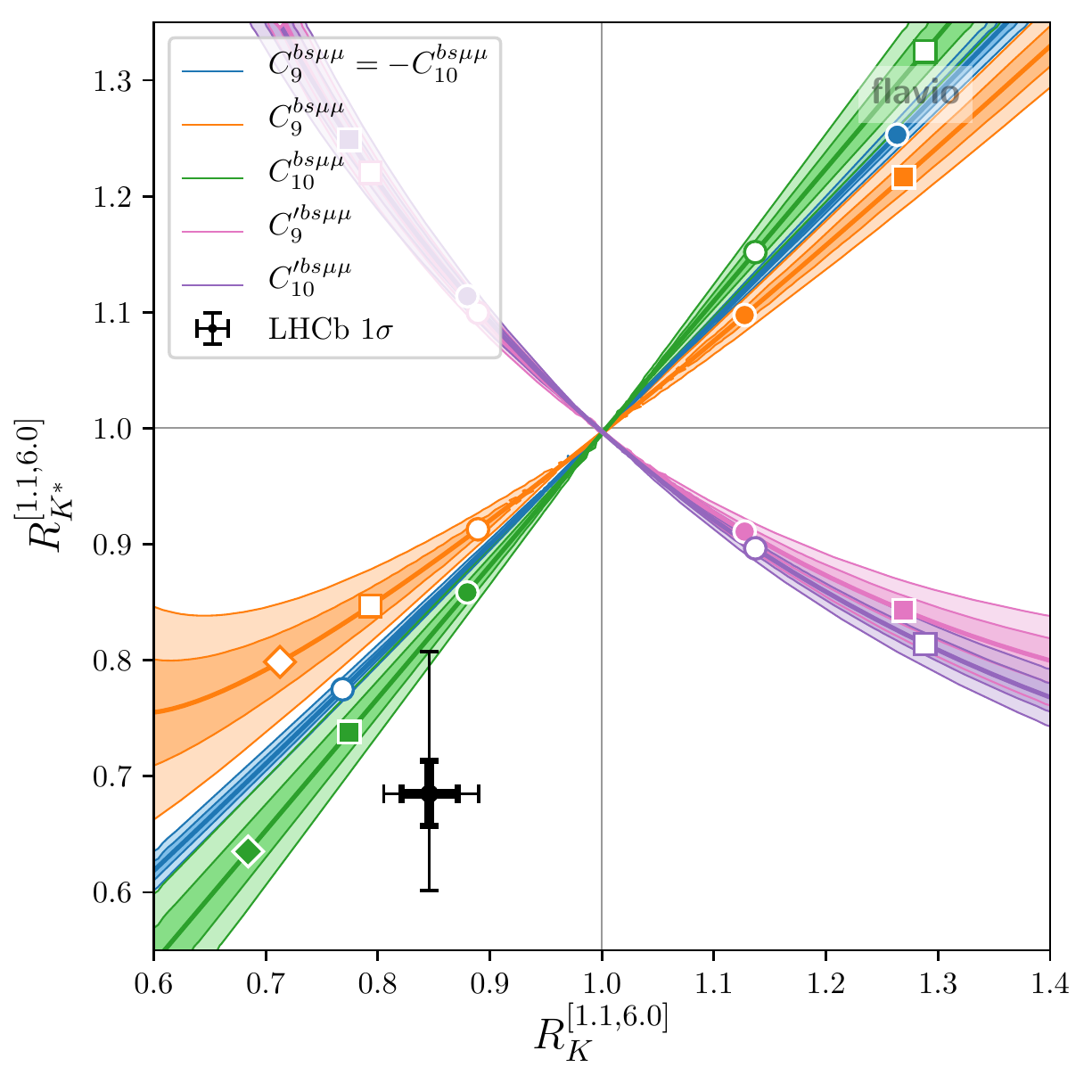}
\includegraphics[width=0.49\textwidth]{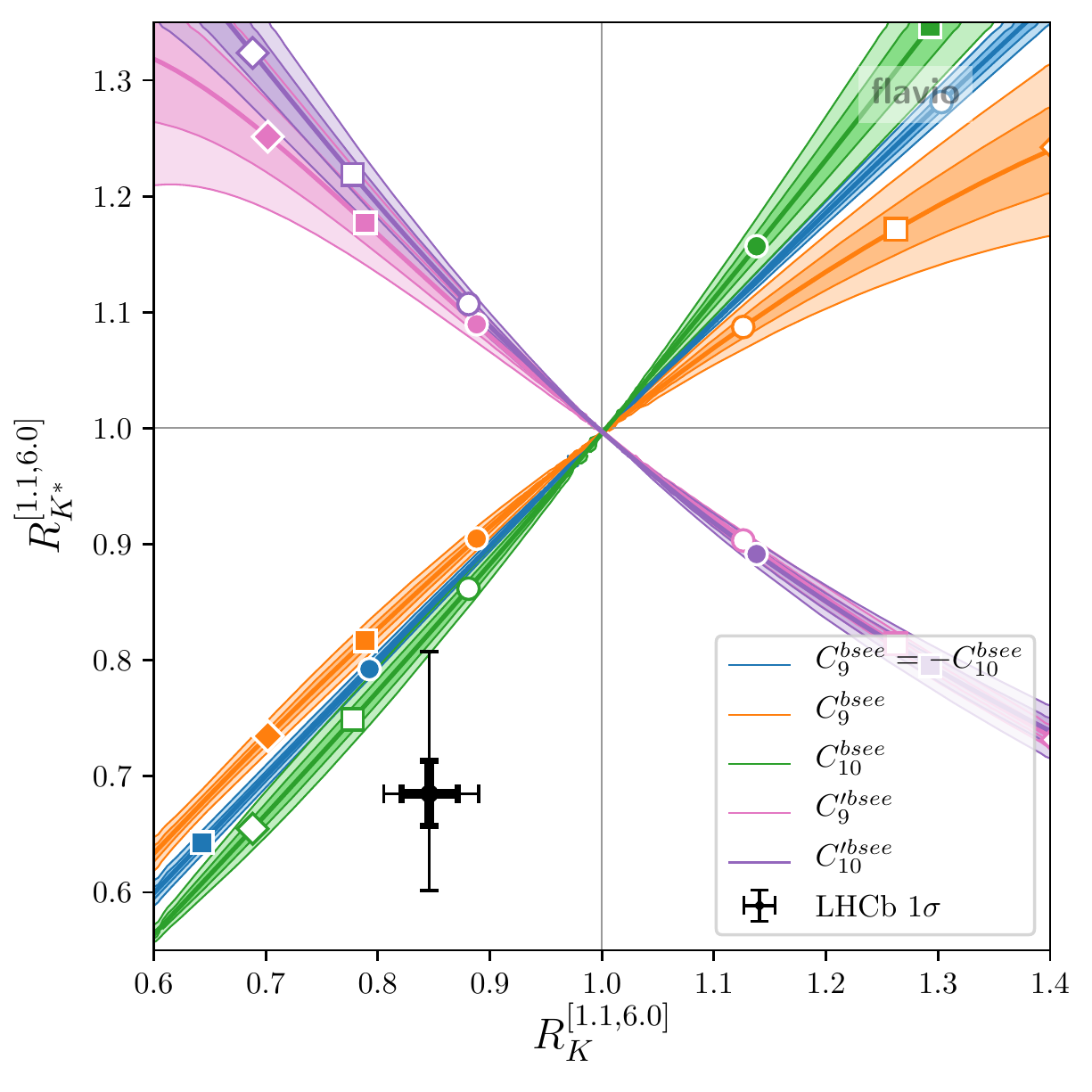}
\caption{Theory predictions for $R_K$ and $R_{K^*}$ in the presence of various non-standard Wilson coefficients (left: new physics in muons; right: new physics in electrons). The colored bands correspond to the $1\sigma$ and $2\sigma$ theory uncertainties.
Circle, square, and diamond markers correspond to Wilson coefficient magnitudes of 0.5, 1.0, and 1.5.
Colored markers correspond to positive, white markers to negative values.
Also shown are the current experimental results (thin error bars) and the expected experimental precision after run 3 of the LHC (bold error bars).}
\label{fig:RKRKstar}
\end{figure}
\begin{figure}[tb]
\centering
\includegraphics[width=0.48\textwidth]{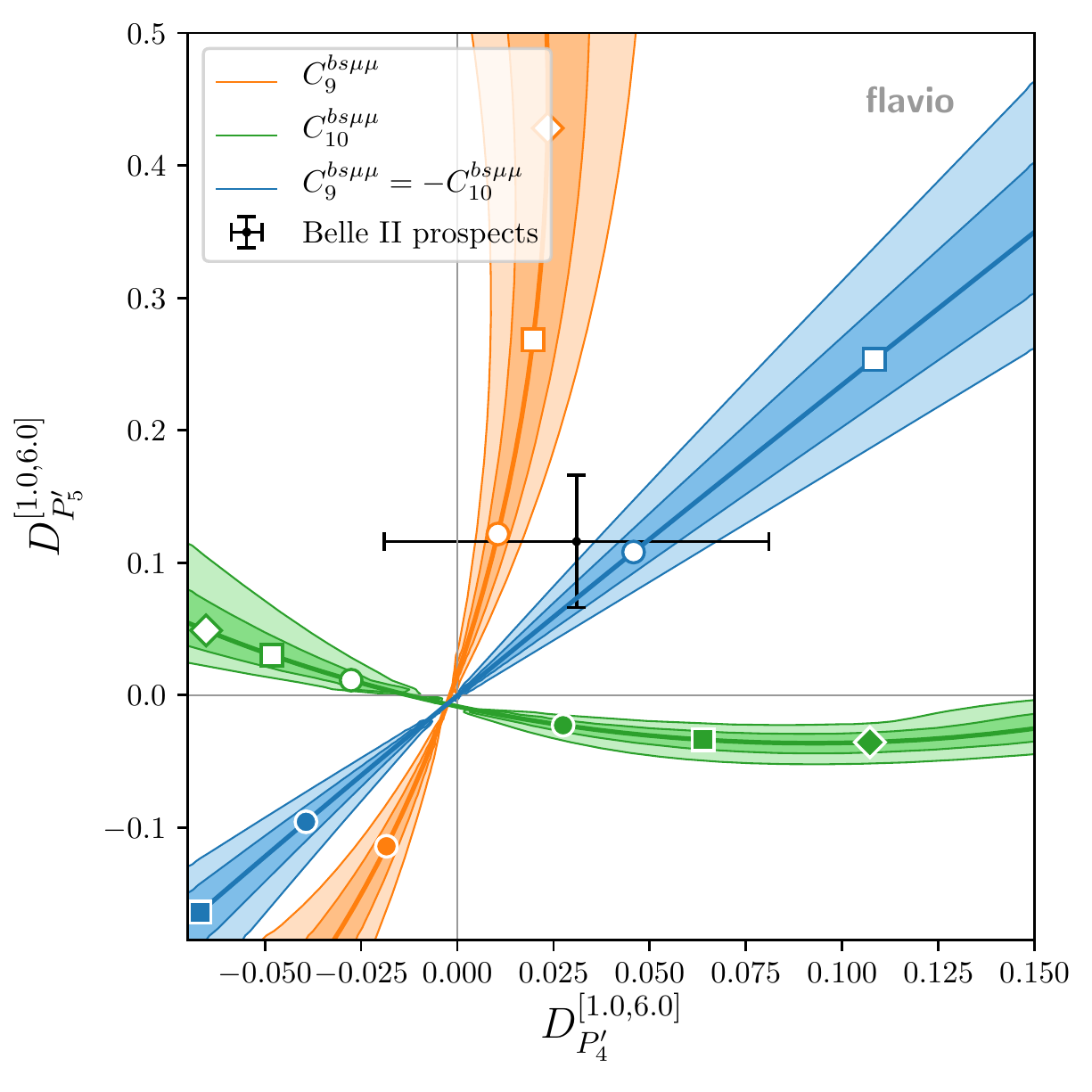}
\caption{Theory predictions for $D_{P_{4}^\prime}$ and $D_{P_{5}^\prime}$ in the presence of few non-standard Wilson coefficients. The colored bands correspond to the $1\sigma$ and $2\sigma$ theory uncertainties.
Circle, square, and diamond markers correspond to Wilson coefficient magnitudes of 0.5, 1.0, and 1.5.
Colored markers correspond to positive, white markers to negative values.
Also shown is the expected experimental precision with the full Belle II data set assuming a new physics benchmark point (black error bars).}
\label{fig:Q4Q5}
\end{figure}

The new error treatment is particularly relevant for quantities that are predicted with very high precision in the SM but that have non-negligible uncertainties in the presence of new physics. In that case, the corresponding entries in the theoretical covariance matrix evaluated in the SM and the ones in the presence of new physics may differ significantly.
The most important examples can be grouped into three categories: (i)~lepton flavor universality tests, (ii)~CP asymmetries, (iii)~observables that vanish in the absence of right-handed currents.
In most cases, the current experimental uncertainties of these observables are considerably larger than the theory uncertainties both in the SM as well as in viable new physics scenarios and the impact of the theory uncertainties in the global fit is moderate. However, with the expected improvement in experimental sensitivity, the theoretical uncertainties will become more and more important and their new physics dependence needs to be taken into account.

Among the lepton universality tests, ratios of branching ratios, like $R_K$ and $R_{K^*}$, are known with high precision in the SM, with uncertainties of around $1\%$~\cite{Bordone:2016gaq, Isidori:2020acz}. In the presence of new physics, however, the uncertainties can be several percent. On the experimental side, the most precisely known quantity is $R_K$, with an uncertainty of $\sim 4\%$~\cite{Aaij:2021vac}, c.f. eq.~(\ref{eq:RK_new}). After run 3 of the LHC, with $\sim 25$~fb$^{-1}$ of integrated luminosity collected by LHCb, one expects an experimental uncertainty of $R_K$ ($R_{K^*}$) of $\sim 2.5\%$ ($2.8\%$)~\cite{Albrecht:2017odf} assuming that systematic uncertainties can be controlled. The precision might reach  $\sim 1\%$ with $300$~fb$^{-1}$.
This clearly shows the need to consistently take into account the theory uncertainties including their new physics dependence.
Other lepton universality tests, like the differences of angular observables $D_{P_{i}^\prime} = P_{i}^\prime(B\to K^* \mu\mu) - P_{i}^\prime(B\to K^* ee)$~\cite{Altmannshofer:2015mqa} (denoted by $Q_i$ in~\cite{Capdevila:2016ivx,Wehle:2016yoi}), have currently sizeable experimental uncertainties~\cite{Wehle:2016yoi} and do not play a major role in global fits, yet. However, given the expected future experimental precision of a few percent~\cite{Albrecht:2017odf} it becomes desirable to have a robust treatment of their theory uncertainties as well.

In Figures~\ref{fig:RKRKstar} and~\ref{fig:Q4Q5} we illustrate the above points with a few examples. The plots in Figure~\ref{fig:RKRKstar} show the theory predictions for $R_K$ and $R_{K^*}$ (in the $q^2$ bin from $1.1$~GeV$^2$ to $6$~GeV$^2$) in the presence of new physics parameterized by various Wilson coefficients. As is well known, the Wilson coefficients with left-handed quark currents ($C_9$ and $C_{10}$) lead to a correlated effect in $R_K$ and $R_{K^*}$, while for right-handed quark currents ($C_9^\prime$ and $C_{10}^\prime$) one finds an anti-correlation~\cite{Hiller:2014ula}. For $C_9 = -C_{10}$ one has to an excellent approximation $R_K \simeq R_{K^*}$. The various colored bands show the theoretical uncertainties at the $1\sigma$ and $2\sigma$ level. Circle, square, and diamond markers correspond to Wilson coefficient magnitudes of 0.5, 1.0, and 1.5.
Colored markers correspond to positive, white markers to negative values.
While the uncertainties are negligible close to the SM point, they become sizeable away from it. For comparison, we also show the current experimental results with $1\sigma$ uncertainties~\cite{Aaij:2017vbb,Aaij:2021vac}, as well as the expected uncertainties after run 3, assuming the same central value.

Similarly, the plot in Figure~\ref{fig:Q4Q5} shows the theory predictions for $D_{P_{4}^\prime}$ and $D_{P_{5}^\prime}$ (in the $q^2$ bin from $1$~GeV$^2$ to $6$~GeV$^2$) in the presence of a few combinations of non standard Wilson coefficients. Also here we observe that the theory uncertainties can be sizable away from the SM point. As the current experimental uncertainties are still large~\cite{Wehle:2016yoi}, we show as comparison the expected experimental uncertainties with the full Belle II data-set which we expect to be around $5\%$\footnote{This value is informed by the expected sensitivities for $P_{4,5}^\prime$ given in~\cite{Albrecht:2017odf} and assumes that $D_{P_{4,5}^\prime}$ can be measured with similar precision.}, assuming as central value the prediction of a new physics benchmark point $(C_9^{bs\mu\mu},C_{10}^{bs\mu\mu}) \simeq (-0.51,0.30)$.

\begin{figure}[tb]
\centering
\includegraphics[width=0.5\textwidth]{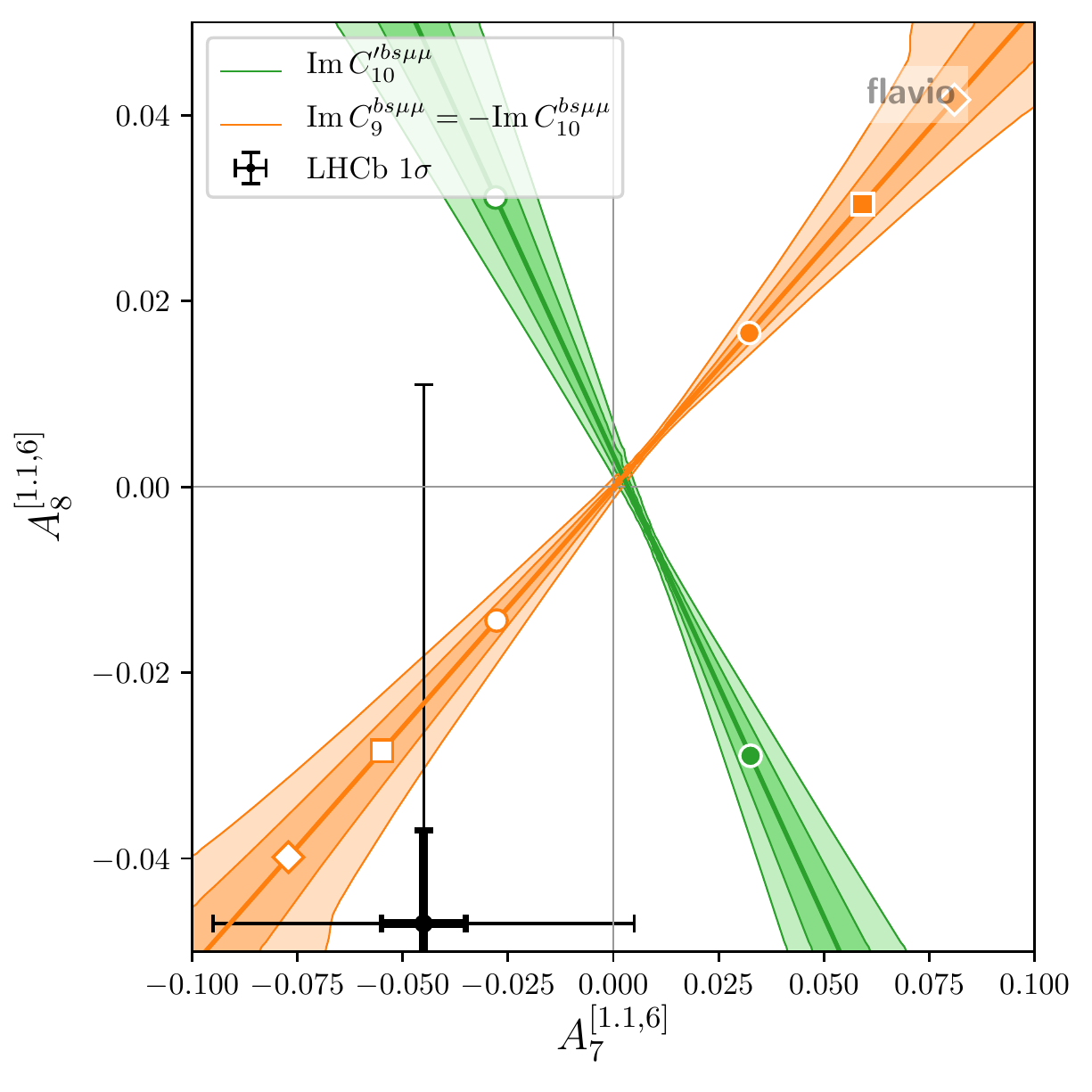}
\caption{Theory predictions for $A_7$ and $A_8$ in the presence of imaginary Wilson coefficients. The colored bands correspond to the $1\sigma$ and $2\sigma$ theory uncertainties.
Circle, square, and diamond markers correspond to Wilson coefficient magnitudes of 0.5, 1.0, and 1.5.
Colored markers correspond to positive, white markers to negative values.
Also shown are the current experimental results (thin error bars) and a experimental precision goal of $1\%$ (bold error bars).}
\label{fig:A7A8}
\end{figure}

With regards to CP violation, we note that results on CP asymmetries in ${B \to K^* \mu^+\mu^-}$ are available from LHCb with $3$fb$^{-1}$ of run 1 data~\cite{Aaij:2015oid}. The most interesting asymmetries are $A_7$, $A_8$, and $A_9$ as they are not suppressed by small strong phases and therefore could in principle be $\mathcal O(1)$ in the presence of CP violating new physics~\cite{Bobeth:2008ij} (Interesting CP asymmetries in $B \to K \mu^+\mu^-$ have been recently discussed in~\cite{Becirevic:2020ssj}). In the SM, they are strongly Cabibbo suppressed, $A_7, A_8 \sim \mathcal{O}(10^{-3})$~\cite{Altmannshofer:2008dz}. The available experimental results are all compatible with zero with uncertainties of approximately $5\%$~\cite{Aaij:2015oid} both at low $q^2 \in (1~\text{GeV}^2, 6~\text{GeV}^2)$ and at high $q^2 \in (15~\text{GeV}^2, 19~\text{GeV}^2)$. Scaling with $\sqrt{N}$, we expect sensitivities with the run 2 data set of approximately $2\% - 3\%$ and ultimate sensitivities of below $1\%$ with $300$fb$^{-1}$.

In Figure~\ref{fig:A7A8} we show the theory predictions for the $B \to K^* \mu^+\mu^-$ CP asymmetries $A_7$ and $A_8$ (in the $q^2$ bin from $1.1$~GeV$^2$ to $6$~GeV$^2$) in the presence of imaginary parts of Wilson coefficients. Similarly to the LFU observables discussed above, also here we observe non-negligible theory uncertainties away from the SM point. For comparison, we also show the current experimental results with $1\sigma$ uncertainties~\cite{Aaij:2015oid}, as well as uncertainties of $1\%$, assuming the same central value.

\section{The Updated Global Fit} \label{sec:fits}

In comparison to our previous fit in~\cite{Aebischer:2019mlg}, we improve the treatment of the theory uncertainties as described in the previous section and we include a series of new experimental results:
\begin{itemize}
 \item The update of the $B^0 \to K^{*0} \mu^+ \mu^-$ angular analysis with 2016 data from LHCb~\cite{Aaij:2020nrf}. The $P_5^\prime$ anomaly persists in this recent update, with a slightly reduced significance compared to the run 1 results~\cite{Aaij:2015oid}. Included in our fit are the angular observables $F_L$, $P_1$, $P_2$, $P_3$, $P_4^\prime$, $P_5^\prime$, $P_6^\prime$, and $P_8^\prime$ in all available $q^2$ bins below $6$\,GeV$^2$ and the one large $q^2$ bin above the narrow charmonium resonances.
 \item The new $B^\pm \to K^{*\pm} \mu^+ \mu^-$ angular analysis~\cite{Aaij:2020ruw}. While the experimental uncertainties of the $B^\pm \to K^{*\pm} \mu^+ \mu^-$ angular analysis are still sizeable, deviations from SM predictions are observed that are broadly showing the same pattern as in the $B^0 \to K^{*0} \mu^+ \mu^-$ angular analysis.
 \item The latest results on $B_s \to \mu^+\mu^-$ from CMS~\cite{Sirunyan:2019xdu} and the very recent result from LHCb~\cite{LHCb:2021awg,LHCb:2021vsc}. We combine these results with the ATLAS result~\cite{Aaboud:2018mst}, as described in appendix~\ref{app:Bsmumu}.  Compared to the previous LHC combination~\cite{LHCb:2020zud}, our combination has a slightly larger central value and a slightly reduced relative uncertainty.
 \item The recent update of $R_K$~\cite{Aaij:2021vac}. The new result has exactly the same central value but reduced uncertainty compared to the previous result~\cite{Aaij:2019wad}, increasing the tension with the SM from $2.5\sigma$ to $3.1\sigma$.
 \item The latest results from LHCb and CMS on the effective $B_s \to \mu^+\mu^-$ lifetime, $\tau_\text{eff} = (2.07\pm 0.29\pm 0.03)\times 10^{-12}$~s~\cite{LHCb:2021awg,LHCb:2021vsc} and $\tau_\text{eff} = (1.70^{+0.61}_{-0.44})\times 10^{-12}$~s~\cite{Sirunyan:2019xdu} (see~\cite{Aaij:2017vad} for the previous LHCb result). Precision measurements of $\tau_\text{eff}$ can lead to non-trivial constraints on new physics in the form of the scalar Wilson coefficients $C_{S,P}^{(\prime)}$~\cite{DeBruyn:2012wk, Altmannshofer:2017wqy}.
 \item The recent update of the $B_s \to \phi \mu^+\mu^-$ branching ratio~\cite{Aaij:2021pkz} that confirms the previously seen tension~\cite{Aaij:2015esa} with the SM prediction.
\end{itemize}

Our numerical code is based on the Python package \texttt{flavio}~\cite{Straub:2018kue}, which provides all the theory predictions including their uncertainties and correlations.
We use the full set of $b\to s\ell\ell$ observables and measurements as implemented in the Python package \texttt{smelli v2.3.1}~\cite{Aebischer:2018iyb,Stangl:2020lbh}, which builds upon \texttt{flavio v2.3.0}.
We plan to implement our new error treatment (cf. section~\ref{sec:uncertainties}) in future versions of \texttt{flavio} and \texttt{smelli}.

\subsection{One parameter scenarios} \label{sec:1parameter}

We start by considering simple one parameter new physics scenarios, switching on one real new physics Wilson coefficient at a time.
We consider several fits, including certain subsets of observables.
In Table~\ref{tab:1d} we report the best fit values for the Wilson coefficients as well as the $1\sigma$ best-fit regions and the ``pull'' in $\sigma$, defined as the $\sqrt{\Delta \chi^2}$ between the best fit point and the $\chi^2$ of the SM.

\afterpage{%
\begin{table}[tbp]
\centering
\renewcommand{\arraystretch}{1.5}
\rowcolors{2}{gray!15}{white}
\addtolength{\tabcolsep}{-1pt} 
\begin{tabularx}{\textwidth}{c|cc|cc|cX}
\toprule
\rowcolor{white}
        & \multicolumn{2}{c|}{$b\to s\mu\mu$} & \multicolumn{2}{c|}{LFU, $B_s\to\mu\mu$} & \multicolumn{2}{c}{all rare $B$ decays}\\
Wilson coefficient  & best fit & pull & best fit & pull & best fit & pull\\
\midrule
$C_9^{bs\mu\mu}$ & $-0.75_{-0.23}^{+0.22}$ & $3.4\sigma$ & $-0.74_{-0.21}^{+0.20}$ & $4.1\sigma$ & $-0.73_{-0.15}^{+0.15}$ & $5.2\sigma$ \\
$C_{10}^{bs\mu\mu}$ & $+0.42_{-0.24}^{+0.23}$ & $1.7\sigma$ & $+0.60_{-0.14}^{+0.14}$ & $4.7\sigma$ & $+0.54_{-0.12}^{+0.12}$ & $4.7\sigma$ \\
$C_9^{\prime bs\mu\mu}$ & $+0.24_{-0.26}^{+0.27}$ & $0.9\sigma$ & $-0.32_{-0.17}^{+0.16}$ & $2.0\sigma$ & $-0.18_{-0.14}^{+0.13}$ & $1.4\sigma$ \\
$C_{10}^{\prime bs\mu\mu}$ & $-0.16_{-0.16}^{+0.16}$ & $1.0\sigma$ & $+0.06_{-0.12}^{+0.12}$ & $0.5\sigma$ & $+0.02_{-0.10}^{+0.10}$ & $0.2\sigma$ \\
$C_9^{bs\mu\mu}=C_{10}^{bs\mu\mu}$ & $-0.20_{-0.15}^{+0.15}$ & $1.3\sigma$ & $+0.43_{-0.18}^{+0.18}$ & $2.4\sigma$ & $+0.05_{-0.12}^{+0.12}$ & $0.4\sigma$ \\
$C_9^{bs\mu\mu}=-C_{10}^{bs\mu\mu}$ & $-0.53_{-0.13}^{+0.13}$ & $3.7\sigma$ & $-0.35_{-0.08}^{+0.08}$ & $4.6\sigma$ & $-0.39_{-0.07}^{+0.07}$ & $5.6\sigma$ \\
\midrule
$C_9^{bsee}$ &  &  & $+0.74_{-0.19}^{+0.20}$ & $4.1\sigma$ & $+0.75_{-0.19}^{+0.20}$ & $4.1\sigma$ \\
$C_{10}^{bsee}$ &  &  & $-0.67_{-0.18}^{+0.17}$ & $4.2\sigma$ & $-0.66_{-0.17}^{+0.17}$ & $4.3\sigma$ \\
$C_9^{\prime bsee}$ &  &  & $+0.36_{-0.17}^{+0.18}$ & $2.1\sigma$ & $+0.40_{-0.18}^{+0.19}$ & $2.3\sigma$ \\
$C_{10}^{\prime bsee}$ &  &  & $-0.31_{-0.16}^{+0.16}$ & $2.1\sigma$ & $-0.30_{-0.16}^{+0.15}$ & $2.0\sigma$ \\
$C_9^{bsee}=C_{10}^{bsee}$ &  &  & $-1.39_{-0.26}^{+0.26}$ & $4.0\sigma$ & $-1.28_{-0.23}^{+0.24}$ & $4.1\sigma$ \\
$C_9^{bsee}=-C_{10}^{bsee}$ &  &  & $+0.37_{-0.10}^{+0.10}$ & $4.2\sigma$ & $+0.37_{-0.10}^{+0.10}$ & $4.3\sigma$ \\
\midrule
$\left(C_S^{bs\mu\mu}=-C_P^{bs\mu\mu}\right)\times\text{GeV}$ &  &  & $-0.004_{-0.002}^{+0.002}$ & $2.1\sigma$ & $-0.003_{-0.002}^{+0.002}$ & $1.4\sigma$ \\
$\left(C_S^{\prime bs\mu\mu}=C_P^{\prime bs\mu\mu}\right)\times\text{GeV}$ &  &  & $-0.004_{-0.002}^{+0.002}$ & $2.1\sigma$ & $-0.003_{-0.002}^{+0.002}$ & $1.4\sigma$    \\

\bottomrule
\end{tabularx}
\addtolength{\tabcolsep}{-4pt} 
\caption{Best-fit values with corresponding $1\sigma$ ranges as well as pulls in sigma between the best-fit point and the SM point for scenarios with NP in a single real Wilson coefficient.
Column ``$b\to s\mu\mu$'': fit including only the $b \to s \mu\mu$ observables (branching ratios and angular observables).
Column ``LFU, $B_s\to\mu\mu$'': fit including only the neutral current LFU observables ($R_{K^{(*)}}$, $D_{P_{4,5}^\prime}$) and BR$(B_s \to \mu^+\mu^-)$.
In column ``all rare $B$ decays'', we show the results of the combined fit.
For the scalar Wilson coefficients, the SM-like solution is shown, while a sign-flipped solution is also allowed~\cite{Altmannshofer:2017wqy}.
}
\label{tab:1d}
\end{table}
\clearpage}

In the column ``$b\to s\mu\mu$'' in Table~\ref{tab:1d}, we focus on the $b \to s \mu\mu$ observables that include the differential branching ratios of $B \to K \mu^+\mu^-$, $B \to K^* \mu^+\mu^-$, $B_s \to \phi \mu^+\mu^-$, and $\Lambda_b \to \Lambda \mu^+\mu^-$ as well as all available CP averaged angular observables in these decays. Note that these observables are subject to potentially large hadronic uncertainties. While existing calculations indicate that long distance effects are well within the assumed uncertainties~\cite{Gubernari:2020eft}, it cannot be fully excluded that such effects are unexpectedly large.
As the considered decay modes do neither involve electrons nor are sensitive to scalar operators, only results for vector and axial-vector muonic Wilson coefficients are shown.
Consistent with previous findings, we observe that a negative $C_9^{bs\mu\mu} \simeq -0.75$ or the left-handed muon combination $C_9^{bs\mu\mu} = - C_{10}^{bs\mu\mu} \simeq -0.53$, are strongly preferred by the fit. For those values of the Wilson coefficients the agreement between theory and data is improved by more than $3\sigma$ compared to the SM

In the column ``LFU, $B_s\to \mu\mu$'' in Table~\ref{tab:1d}, we consider the neutral current LFU observables ($R_{K^{(*)}}$, $D_{P_{4,5}^\prime}$) and BR$(B_s \to \mu^+\mu^-)$ only, including in particular the new $R_K$ and BR$(B_s \to \mu^+\mu^-)$ result. The included observables are considered under excellent theoretical control and the discrepancies cannot be explained by hadronic effects. Two scenarios stand out, $C_{10}^{bs\mu\mu} \simeq +0.60$ and $C_{9}^{bs\mu\mu} = - C_{10}^{bs\mu\mu} \simeq -0.35$, which have a pull of $4.7\sigma$ and $4.6\sigma$, respectively. These scenarios do not only address the anomalies in $R_K$ and $R_{K^*}$, but also the slightly reduced branching ratio of $B_s \to \mu^+\mu^-$. The coefficients $C_9^{bs\mu\mu}$, $C_9^{bsee}$, and $C_{10}^{bsee}$ can explain the $R_K$ and $R_{K^*}$ data, but do not affect the $B_s \to \mu^+\mu^-$ decay. Their pulls are therefore a bit lower, around $4\sigma$.
The scalar Wilson coefficients show a slight ($\sim 2\sigma$) preference for negative values, that lead to a suppression of the $B_s\to \mu\mu$ branching ratio in accordance with the data. Note that we include the effect of the scalar Wilson coefficients only in the $B_s \to \mu^+\mu^-$ decay. In the parameter space allowed by $B_s \to \mu^+\mu^-$, the scalar Wilson coefficients have negligible impact on all the other $b \to s \mu \mu$ transitions.

Finally, in the the column ``all rare $B$ decays'' in Table~\ref{tab:1d} we show the results of the global fit. Included are the $b\to s\mu\mu$ observables, the LFU observables, and the $B_s \to \mu^+\mu^-$ branching ratio.\footnote{%
Note that in previous fits~\cite{Aebischer:2019mlg} we had also included $\Delta F=2$ observables that are correlated to the $B_s \to \mu^+\mu^-$ branching ratio and the various $b \to s \mu\mu$ branching ratios, mainly through their dependence on common CKM input. Adding $\Delta F=2$ observables in the fit further increases the pulls slightly.}
The largest pulls of $5.6\sigma$ and $5.2\sigma$ are found for $C_{9}^{bs\mu\mu} = - C_{10}^{bs\mu\mu} \simeq -0.39$ and $C_{9}^{bs\mu\mu} \simeq -0.73$, respectively. As expected, the pulls for the electronic Wilson coefficients are very similar to the values in the ``LFU, $B_s\to \mu\mu$'' column. We observe a small change in the preferred values for the scalar Wilson coefficients, which is due to the correlations of the theory uncertainties of BR$(B_s \to \mu^+\mu^-)$ and the $b \to s \mu\mu$ observables.

\begin{table}[tbp]
\centering
\renewcommand{\arraystretch}{1.5}
\rowcolors{2}{gray!15}{white}
\addtolength{\tabcolsep}{1.7pt} 
\begin{tabularx}{\textwidth}{cc|cc|cc|cX}
\toprule
\rowcolor{white}
&        & \multicolumn{2}{c|}{$b\to s\mu\mu$} & \multicolumn{2}{c|}{LFU, $B_s\to\mu\mu$} & \multicolumn{2}{c}{all rare $B$ decays}\\
& Wilson coefficient  & best fit & pull & best fit & pull & best fit & pull\\
\midrule
\cellcolor{white}& $C_9^{bs\mu\mu}$ & $-0.75_{-0.23}^{+0.22}$ & $3.4\sigma$ & $-0.74_{-0.21}^{+0.20}$ & $4.1\sigma$ & $-0.73_{-0.15}^{+0.15}$ & $5.2\sigma$ \\
\cellcolor{white}& $C_{10}^{bs\mu\mu}$ & $+0.42_{-0.24}^{+0.23}$ & $1.7\sigma$ & $+0.60_{-0.14}^{+0.14}$ & $4.7\sigma$ & $+0.54_{-0.12}^{+0.12}$ & $4.7\sigma$ \\
\multirow{-3}{*}{\cellcolor{white}\rotatebox{90}{NP errors}} & $C_9^{bs\mu\mu}=-C_{10}^{bs\mu\mu}$ & $-0.53_{-0.13}^{+0.13}$ & $3.7\sigma$ & $-0.35_{-0.08}^{+0.08}$ & $4.6\sigma$ & $-0.39_{-0.07}^{+0.07}$ & $5.6\sigma$ \\
\midrule
\cellcolor{white}& $C_9^{bs\mu\mu}$ & $-0.88_{-0.21}^{+0.22}$ & $3.7\sigma$ & $-0.74_{-0.21}^{+0.20}$ & $4.1\sigma$ & $-0.78_{-0.15}^{+0.15}$ & $5.3\sigma$ \\
\cellcolor{white}& $C_{10}^{bs\mu\mu}$ & $+0.44_{-0.21}^{+0.21}$ & $2.1\sigma$ & $+0.60_{-0.14}^{+0.14}$ & $4.7\sigma$ & $+0.54_{-0.12}^{+0.12}$ & $4.8\sigma$ \\
\multirow{-3}{*}{\cellcolor{white}\rotatebox{90}{SM errors}} & $C_9^{bs\mu\mu}=-C_{10}^{bs\mu\mu}$ & $-0.58_{-0.18}^{+0.17}$ & $3.6\sigma$ & $-0.35_{-0.08}^{+0.08}$ & $4.6\sigma$ & $-0.39_{-0.07}^{+0.07}$ & $5.5\sigma$ \\

\bottomrule
\end{tabularx}
\addtolength{\tabcolsep}{-4pt} 
\caption{Best-fit ranges for selected Wilson coefficients, taking into account the dependence of the theory errors on the Wilson coefficients (first 3 rows) and fixing the theory errors to the SM values (last 3 rows).
}
\label{tab:1d_comp}
\end{table}

To illustrate the impact of our improved treatment of theory uncertainties, we compare in Table~\ref{tab:1d_comp} the fit results in the $C_9^{bs\mu\mu}$, $C_{10}^{bs\mu\mu}$, and $C_9^{bs\mu\mu} = -C_{10}^{bs\mu\mu}$ scenarios taking into account the dependence of the theory errors on the Wilson coefficients (first 3 rows) and fixing the theory errors to the SM values (last 3 rows). We find that the impact is currently still moderate. The largest shift is observed in the $C_9^{bs\mu\mu}$ scenario, in which the pull from the $b \to s \mu\mu$ observables is somewhat reduced once the new physics dependence of the theory errors is taken into account. We expect the effect to become much more pronounced with more precise data.

\subsection{Two parameter scenarios}

\begin{figure}[tb]
\centering
\includegraphics[width=0.49\textwidth]{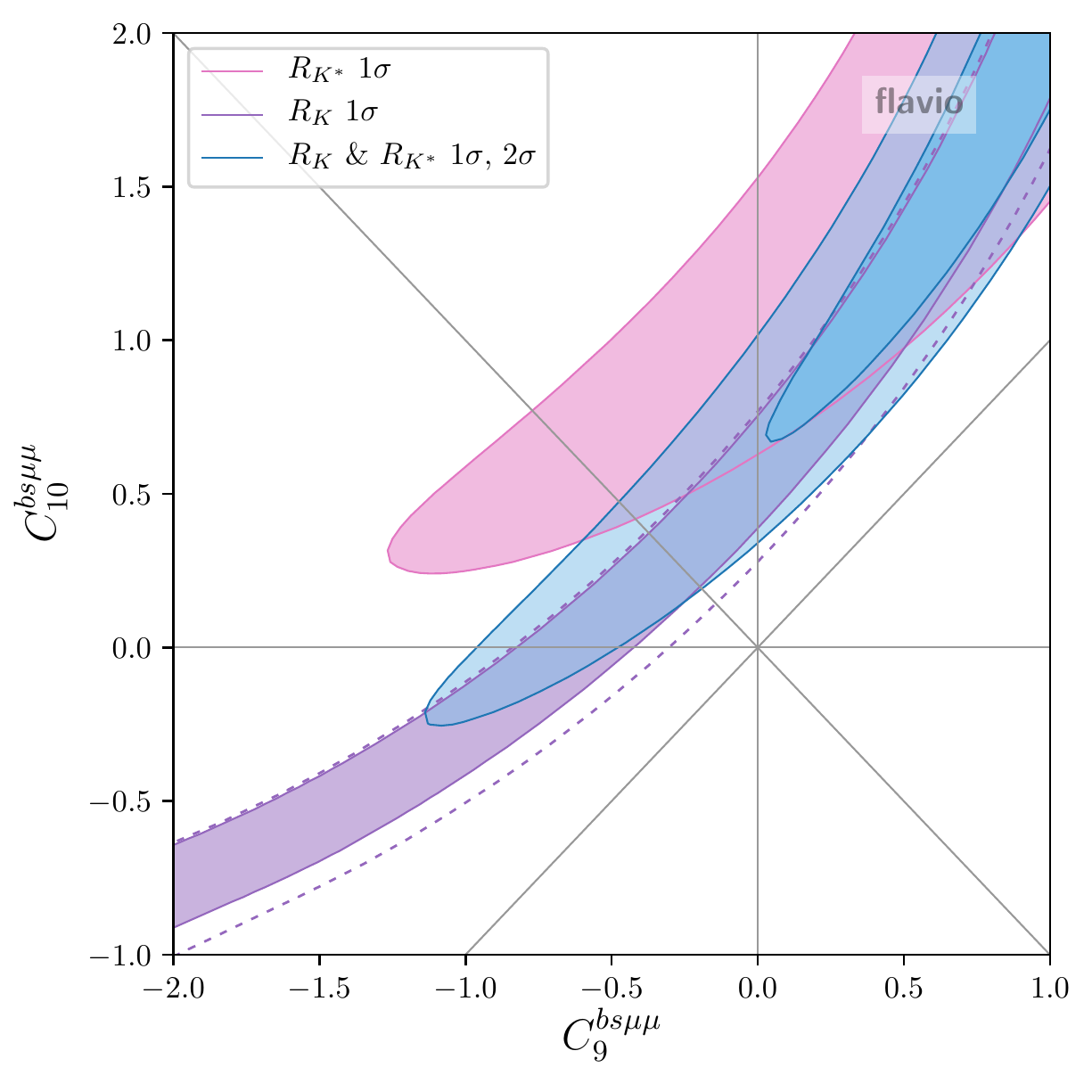}
\includegraphics[width=0.49\textwidth]{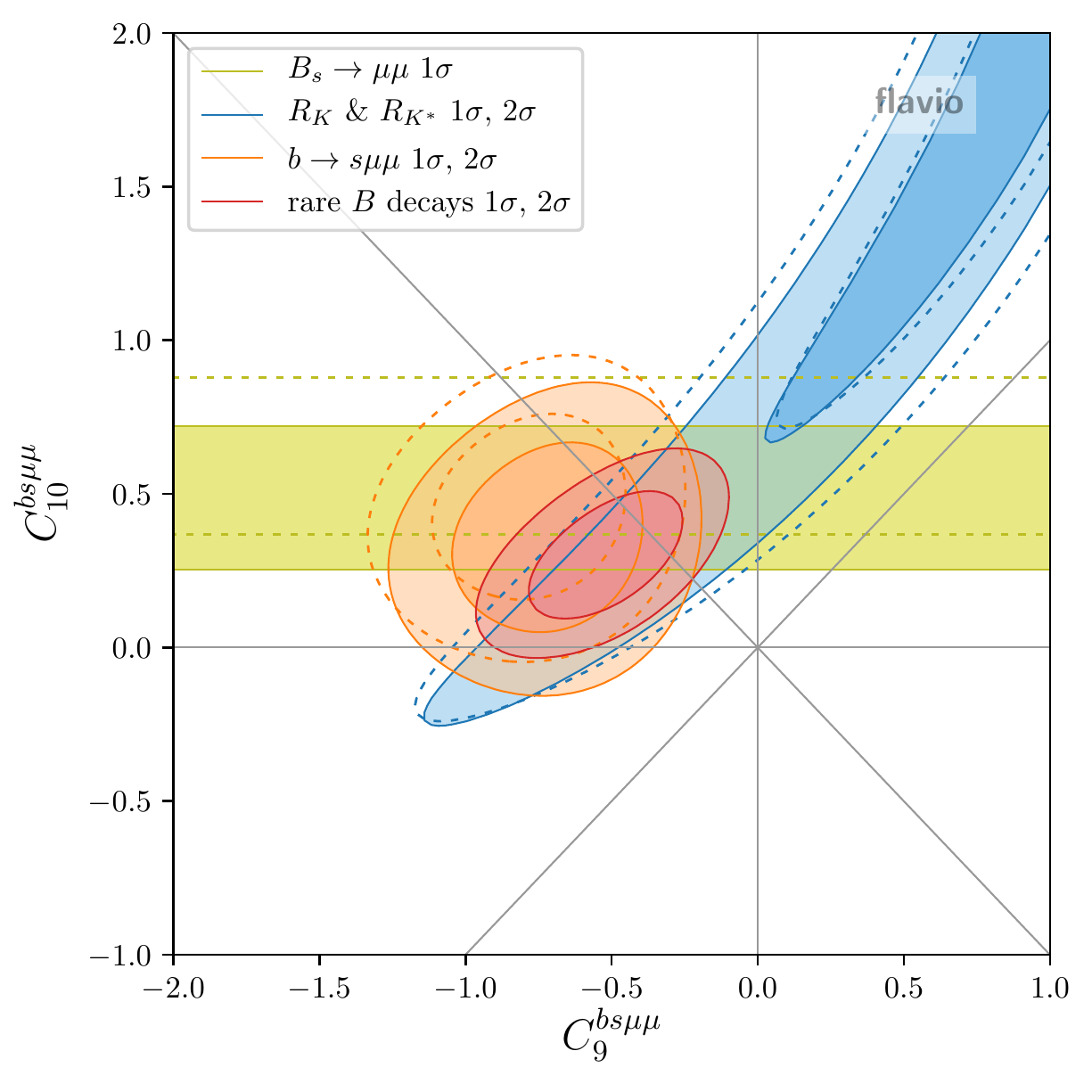}
\caption{Constraints in the Wilson coefficient plane $C_9^{bs\mu\mu}$ vs. $C_{10}^{bs\mu\mu}$. Left: LFU ratios only. Right: Combination of LFU ratios, combination of $b \to s \mu\mu$ observables, BR$(B_s \to \mu^+\mu^-)$, and the global fit. The dashed lines show the constraints before the recent updates~\cite{Aaij:2021vac,LHCb:2021awg,LHCb:2021vsc,Aaij:2021pkz}.}
\label{fig:2d1}
\end{figure}

Next, we discuss scenarios where two Wilson coefficients are turned on simultaneously.
In Figure~\ref{fig:2d1} we show the best fit regions in the $C_9^{bs\mu\mu}$ vs. $C_{10}^{bs\mu\mu}$ plane. The plot on the left focuses on the constraints from the LFU ratios $R_{K}$ and $R_{K^*}$. The $R_K$ constraint before the update~\cite{Aaij:2021vac} is shown by the dashed contours. As the measured $R_K > R_{K^*}$ the best fit range prefers a sizable positive $C_{10}^{bs\mu\mu}$. The plot on the right shows the result of the global fit. The $B_s \to \mu^+\mu^-$ branching ratio prefers a modest positive $C_{10}^{bs\mu\mu}$, while the $b\to s\mu\mu$ observables mainly prefer a negative $C_9^{bs\mu\mu}$. Overall, the best fit point corresponds to $(C_9^{bs\mu\mu},C_{10}^{bs\mu\mu}) \simeq (-0.51,0.30)$ with a pull of $5.3\sigma$.

\begin{figure}[tb]
\centering
\includegraphics[width=0.49\textwidth]{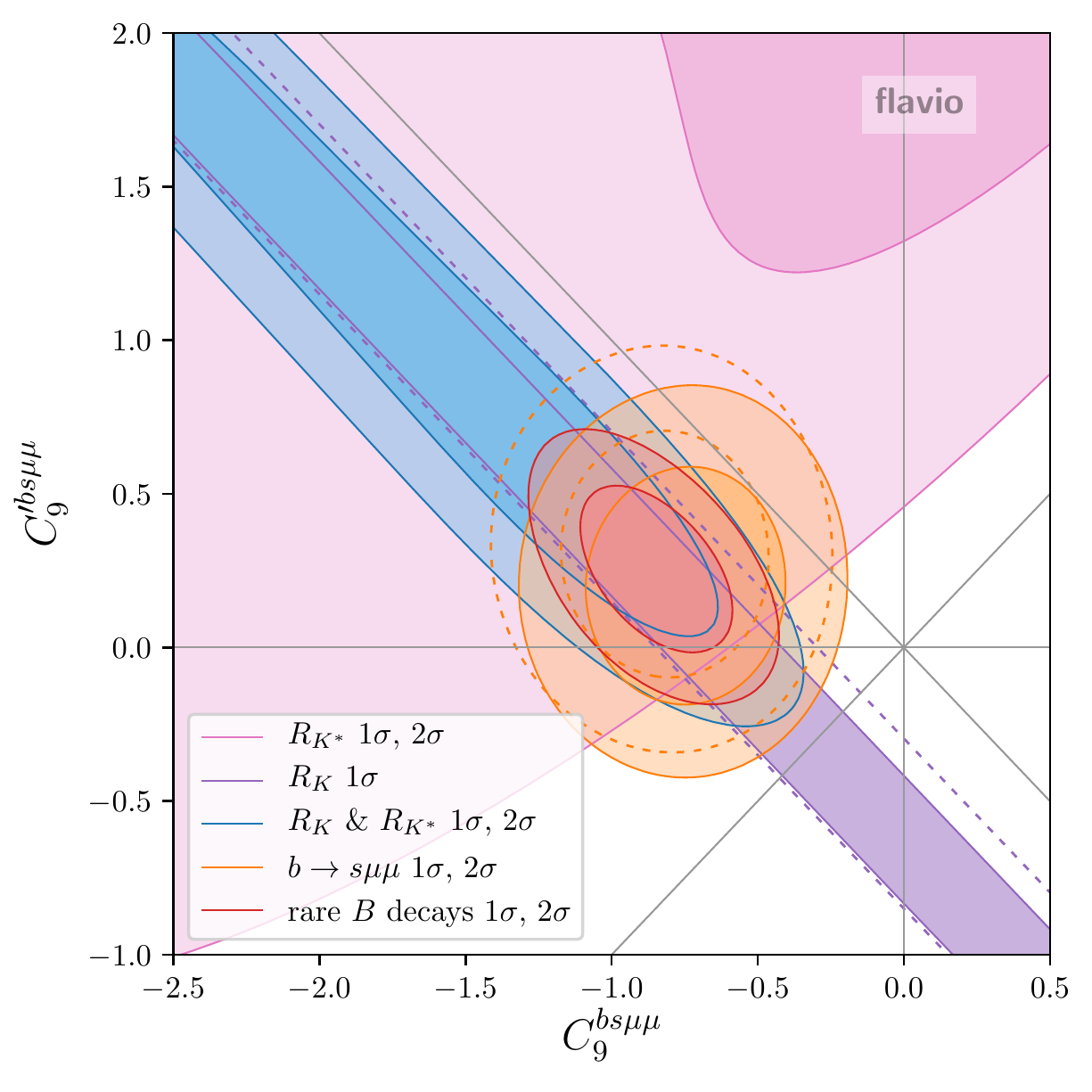}
\includegraphics[width=0.49\textwidth]{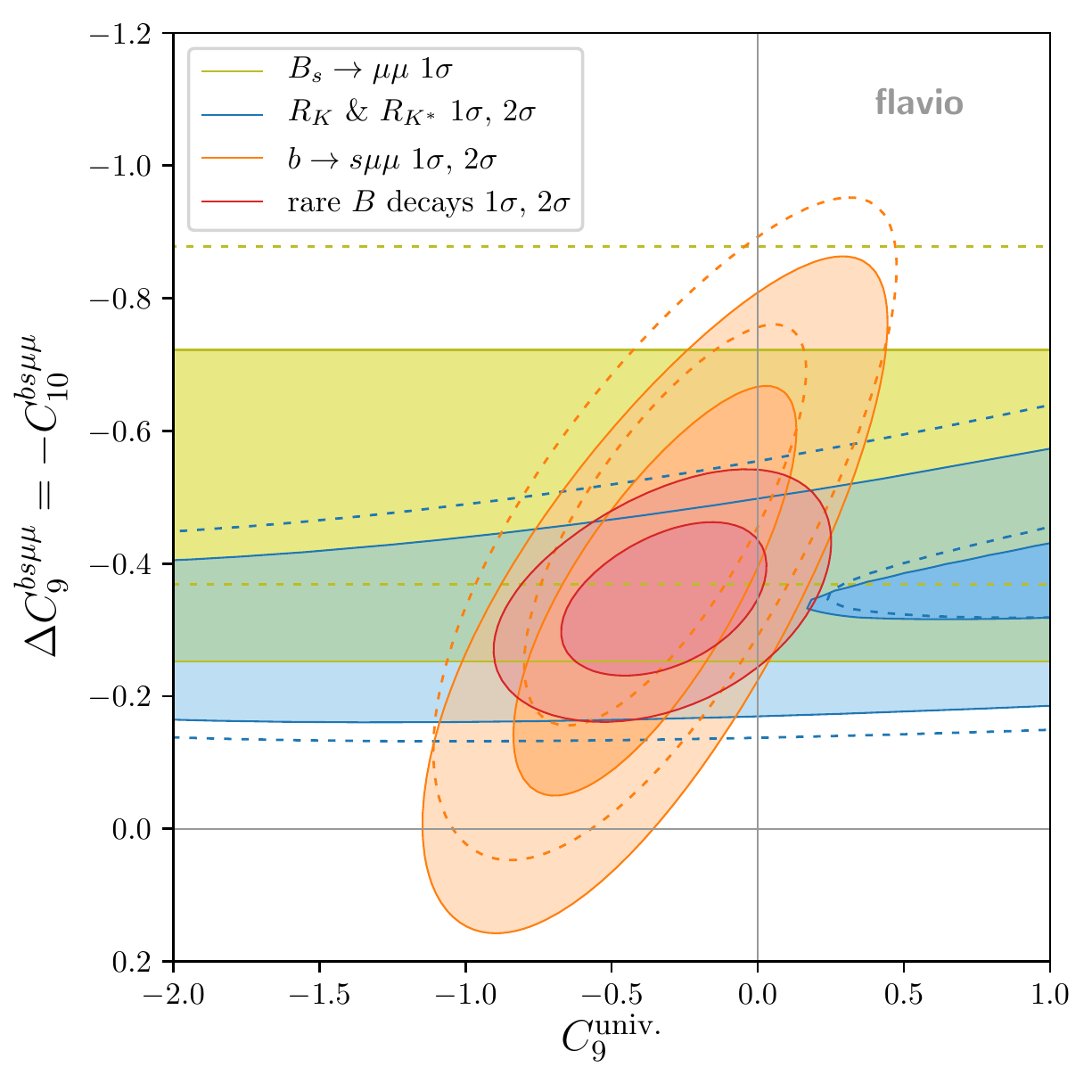}
\caption{Constraints in the Wilson coefficient planes $C_9^{bs\mu\mu}$ vs. $C_{9}^{\prime\,bs\mu\mu}$ (left) and $C_9^\text{univ.}$ vs. $\Delta C_{9}^{bs\mu\mu} = -C_{10}^{bs\mu\mu}$ (right). The dashed lines show the constraints before the recent updates~\cite{Aaij:2021vac,LHCb:2021awg,LHCb:2021vsc,Aaij:2021pkz}.}
\label{fig:2d2}
\end{figure}

In Figure~\ref{fig:2d2} we show the viable parameter space of a couple of other Wilson coefficient pairs, that were found to give good fits in the past. The plot on the left shows the $C_9^{bs\mu\mu}$ vs. $C_{9}^{\prime\,bs\mu\mu}$ plane, while the plot on the right shows the $C_9^\text{univ.}$ vs. $\Delta C_9^{bs\mu\mu} = -C_{10}^{bs\mu\mu}$ plane (defined such that $C_9^{bsee} = C_9^\text{univ.}$ and $C_9^{bs\mu\mu} = C_9^\text{univ.} + \Delta C_9^{bs\mu\mu}$). The best fit points are given by $(C_9^{bs\mu\mu},C_{9}^{\prime\,bs\mu\mu}) \simeq (-0.84,0.25)$ and $(C_9^\text{univ.},\Delta C_{9}^{bs\mu\mu}) \simeq (-0.32,-0.34)$ and correspond to pulls of $5.0\sigma$ and $5.4\sigma$, respectively.
The scenario on the left gives an excellent fit of $R_K$ and $R_{K^*}$, but the slightly reduced $B_s \to \mu^+\mu^-$ branching ratio remains unexplained. The scenario on the right can resolve the tension in BR$(B_s \to \mu^+\mu^-)$, but leaves a tension between $R_K$ and $R_{K^*}$. Note that $C_9^\text{univ.}$ could in principle be mimicked by a hadronic effect. A lepton flavor universal $C_9^\text{univ.}$ of the preferred size can also be generated through renormalization group running from semi-tauonic operators that are motivated by the $R_{D^{(*)}}$ anomalies~\cite{Crivellin:2018yvo} or from four-quark operators~\cite{Aebischer:2019mlg}.

\begin{figure}[tb]
\centering
\includegraphics[width=0.5\textwidth]{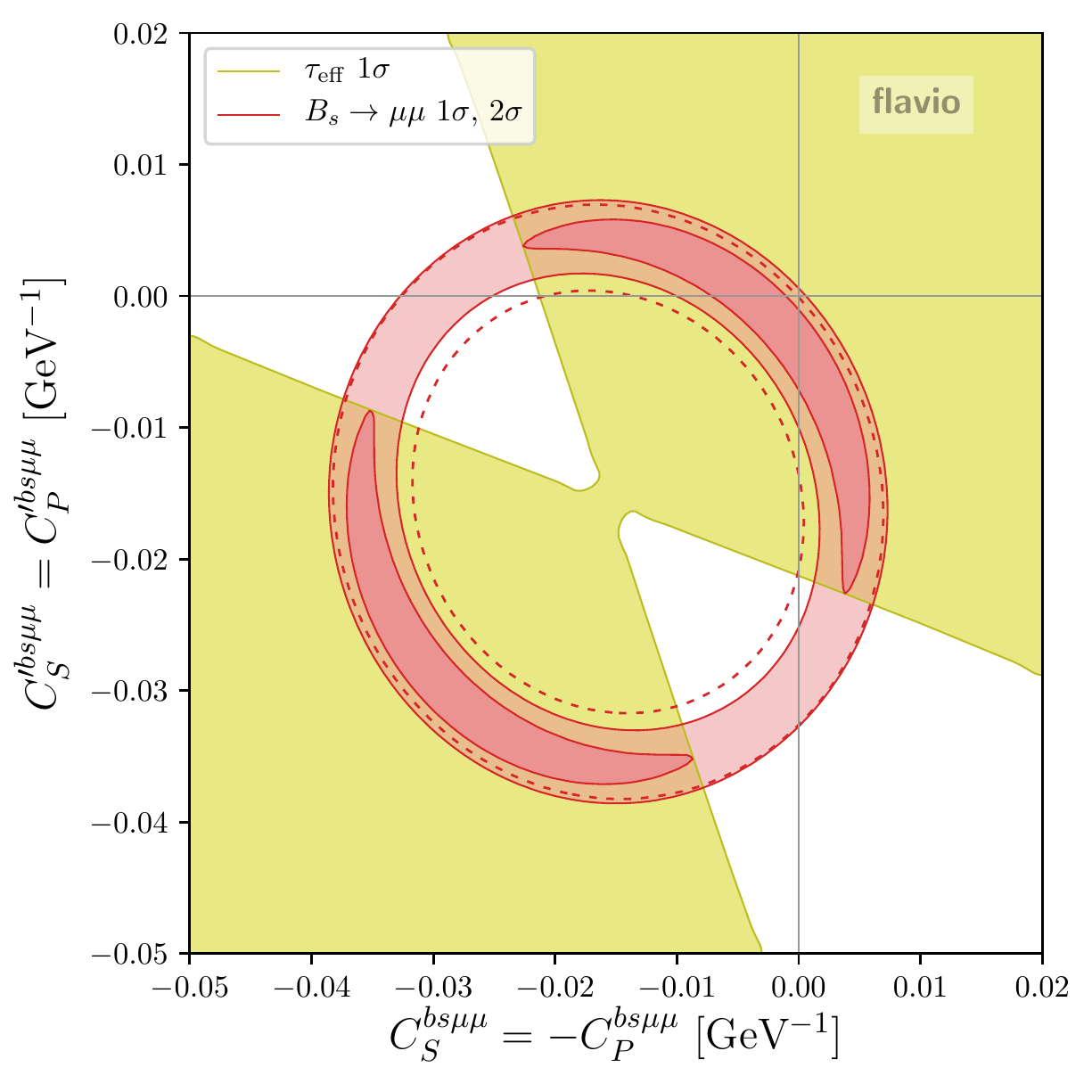}
\caption{Constraint in the Wilson coefficient plane $C_S^{bs\mu\mu} = -C_P^{bs\mu\mu}$ vs. $C_S^{\prime\,bs\mu\mu} = C_P^{\prime\,bs\mu\mu}$. The red band shows at $1\sigma$ and $2\sigma$ the constraints for $(C_{10}^{bs\mu\mu},C_{10}^{\prime\,bs\mu\mu}) = (0,0)$. The dashed lines show the constraints before the recent update~\cite{LHCb:2021awg,LHCb:2021vsc}.}
\label{fig:scalar}
\end{figure}

As clearly seen in the plots of Figures~\ref{fig:2d1} and~\ref{fig:2d2}, the branching ratio of $B_s \to \mu^+\mu^-$ plays an important role in constraining the Wilson coefficient $C_{10}$. It is well known that $B_s \to \mu^+\mu^-$ is also very sensitive to new physics in the scalar Wilson coefficients (see e.g.~\cite{Altmannshofer:2017wqy}). In Figure~\ref{fig:scalar} we show the constraints in the Wilson coefficient plane $C_S^{bs\mu\mu} = -C_P^{bs\mu\mu}$ vs. $C_S^{\prime\,bs\mu\mu} = C_P^{\prime\,bs\mu\mu}$ based on our combination of the experimental results on BR$(B_s \to \mu^+\mu^-)$. Also the available results on the effective $B_s \to \mu^+\mu^-$ lifetime are included in the fit. The red band shows the $1\sigma$ and $2\sigma$ constraint when the semileptonic new physics coefficients $C_{10}^{bs\mu\mu}$ and $C_{10}^{\prime\,bs\mu\mu}$ are set to zero. The $\sim 2\sigma$ tension between the BR$(B_s \to \mu^+\mu^-)$ SM prediction and the experimental world average is clearly reflected in the plot. With the recent BR$(B_s \to \mu^+\mu^-)$ update, the preferred region in the Wilson coefficient space moved slightly towards the SM point.
We observe that the measurements of the effective $B_s \to \mu^+\mu^-$ lifetime already have some impact on the allowed parameter space of the scalar Wilson coefficients. The region of parameter space that corresponds to a mass eigenstate rate asymmetry $A_{\Delta \Gamma} = -1$ is excluded at the $1\sigma$ level.
Note that the latest LHCb result for BR$(B_s \to \mu^+\mu^-)$ assumes the SM value $A_{\Delta \Gamma} = +1$. Due to the lifetime dependence of the acceptance, the experimentally determined BR$(B_s \to \mu^+\mu^-)$ is larger by approximately $5\%$ or $11\%$ for $A_{\Delta \Gamma} = 0$ or $-1$, respectively~\cite{LHCb:2021awg,LHCb:2021vsc}. A similar effect is observed in the ATLAS and CMS analyses~\cite{LHCb:2020zud}. We do not attempt to model this effect in our fit of the scalar Wilson coefficients. In the region that is currently slightly disfavored by the measured effective $B_s \to \mu^+\mu^-$ lifetime, we expect a few percent shift of the best fit band.

\subsection{Generic Scenarios}

We also consider more generic scenarios with more then two Wilson coefficients. In particular, we consider a four parameter scenario including the muon-specific semi-leptonic Wilson coefficients $C_9^{bs\mu\mu}$, $C_{10}^{bs\mu\mu}$, $C_9^{\prime bs\mu\mu}$, and $C_{10}^{\prime bs\mu\mu}$, as well as a six parameter scenario including both muon-specific and electron-specific Wilson coefficients $C_9^{bs\mu\mu}$, $C_{10}^{bs\mu\mu}$, $C_9^{\prime bs\mu\mu}$, $C_{10}^{\prime bs\mu\mu}$, $C_9^{bsee}$, and $C_{10}^{bsee}$.

\begin{table}[tbh]
\centering
\begin{tabular}{l|r}
\multicolumn{2}{c}{$b\to s\mu\mu$}\\
\hline
$C_9^{bs\mu\mu}$ & $-0.84 \pm 0.23$ \\
$C_{10}^{bs\mu\mu}$ & $+0.24 \pm 0.21$ \\
$C_9^{\prime bs\mu\mu}$ & $-0.21 \pm 0.34$ \\
$C_{10}^{\prime bs\mu\mu}$ & $-0.33 \pm 0.22$ \\
\end{tabular}\qquad
\begin{tabular}{l|rrrr}
& $C_9^{bs\mu\mu}$ & $C_{10}^{bs\mu\mu}$ & $C_9^{\prime bs\mu\mu}$ & $C_{10}^{\prime bs\mu\mu}$ \\
\hline
$C_9^{bs\mu\mu}$ & $1\phantom{.00}$ & $0.24$ & $0.37$ & $0.41$ \\
$C_{10}^{bs\mu\mu}$ &  & $1\phantom{.00}$ & $0.13$ & $0.33$ \\
$C_9^{\prime bs\mu\mu}$ &  &  & $1\phantom{.00}$ & $0.71$ \\
$C_{10}^{\prime bs\mu\mu}$ &  &  &  & $1\phantom{.00}$ \\
\end{tabular}
\caption{Best fit values, uncertainties, and correlation matrix of the four-parameter fit to the Wilson coefficients $C_9^{bs\mu\mu}$, $C_{10}^{bs\mu\mu}$, $C_9^{\prime bs\mu\mu}$, and $C_{10}^{\prime bs\mu\mu}$ including only $b\to s\mu\mu$ observables.}
\label{tab:4parameter1}
\end{table}

\begin{table}[tbh]
\centering
\begin{tabular}{l|r}
\multicolumn{2}{c}{all rare $B$ decays}\\
\hline
$C_9^{bs\mu\mu}$ & $-0.83 \pm 0.23$ \\
$C_{10}^{bs\mu\mu}$ & $+0.17 \pm 0.15$ \\
$C_9^{\prime bs\mu\mu}$ & $-0.08 \pm 0.30$ \\
$C_{10}^{\prime bs\mu\mu}$ & $-0.33 \pm 0.19$ \\
\end{tabular}\qquad
\begin{tabular}{l|rrrr}
& $C_9^{bs\mu\mu}$ & $C_{10}^{bs\mu\mu}$ & $C_9^{\prime bs\mu\mu}$ & $C_{10}^{\prime bs\mu\mu}$ \\
\hline
$C_9^{bs\mu\mu}$ & $1\phantom{.00}$ & $0.66$ & $0.38$ & $0.58$ \\
$C_{10}^{bs\mu\mu}$ &  & $1\phantom{.00}$ & $0.54$ & $0.55$ \\
$C_9^{\prime bs\mu\mu}$ &  &  & $1\phantom{.00}$ & $0.81$ \\
$C_{10}^{\prime bs\mu\mu}$ &  &  &  & $1\phantom{.00}$ \\
\end{tabular}
\caption{Best fit values, uncertainties, and correlation matrix of the four-parameter fit to the Wilson coefficients $C_9^{bs\mu\mu}$, $C_{10}^{bs\mu\mu}$, $C_9^{\prime bs\mu\mu}$, and $C_{10}^{\prime bs\mu\mu}$ including all observables.}
\label{tab:4parameter2}
\end{table}

In the four parameter scenario we perform two fits: (1) a fit including only the $b \to s \mu\mu$ observables (branching ratios and CP averaged angular observables) and (2) the global fit of all rare B decay data, including the LFU observables and $B_s \to \mu^+\mu^-$. In both cases we identify the best fit point in Wilson coefficient space and approximate the likelihood function in its vicinity by a multivariate Gaussian. The parameters of the multivariate Gaussians (i.e. the central values for the Wilson coefficients, their uncertainties and the correlation matrix) are determined by the \texttt{migrad} and \texttt{hesse} algorithms implemented in the \texttt{iminuit}~\cite{iminuit,James:1975dr} Python package. The corresponding values are given in Tables~\ref{tab:4parameter1} and~\ref{tab:4parameter2}. The results for the central values agree well within the uncertainties and we observe slightly smaller uncertainties in the global fit. The fits prefer new physics in $C_9^{bs\mu\mu}$ with large significance. The corresponding central value is close to the result found in the one-parameter fit to $C_9^{bs\mu\mu}$ discussed in section~\ref{sec:1parameter}.

We find sizable correlations among the Wilson coefficients. One of the main contributors to the correlations is the new precise measurement of $R_K$, as can be seen in the two parameter scenarios shown in Figures~\ref{fig:2d1} and~\ref{fig:2d2}. We find a positive correlation between $C_9^{bs\mu\mu}$ and $C_{10}^{bs\mu\mu}$ that increases when $R_K$ is included, as expected from Figure~\ref{fig:2d1}. In the four parameter fit, we also find a sizable positive correlation between $C_9^{bs\mu\mu}$ and $C_9^{\prime bs\mu\mu}$, and large positive correlations between $C_{10}^{\prime bs\mu\mu}$ and the other Wilson coefficients. From Figure~\ref{fig:2d2} one might expect a negative correlation between $C_9^{bs\mu\mu}$ and $C_9^{\prime bs\mu\mu}$. We find that this is indeed the case for fixed values of $C_{10}^{\prime bs\mu\mu}$. However, the large correlations of $C_{10}^{\prime bs\mu\mu}$ lead to an overall positive correlation when the four dimensional likelihood is projected onto the $C_9^{bs\mu\mu}$ - $C_9^{\prime bs\mu\mu}$ plane.

The central values of our global four parameter fit agree within uncertainties with the central values of a similar fit performed in~\cite{1853015}. Compared to~\cite{1853015} we find a much larger positive correlation between $C_{10}^{\prime bs\mu\mu}$ and the other Wilson coefficients.
This leads in our fit to a slightly negative central value for $C_{9}^{\prime bs\mu\mu}$ and a positive correlation between $C_9^{bs\mu\mu}$ and $C_{9}^{\prime bs\mu\mu}$ compared to a positive central value for $C_{9}^{\prime bs\mu\mu}$ and slightly negative correlation in~\cite{1853015}. We checked that excluding the high-$q^2$ bins from our fit (as done in~\cite{1853015}) improves the agreement with~\cite{1853015} to some extent, but differences remain.

\begin{table}[tb]
\centering
\begin{tabular}{l|r}
\multicolumn{2}{c}{all rare $B$ decays}\\
\hline
$C_9^{bs\mu\mu}$ & $-0.82 \pm 0.23$ \\
$C_{10}^{bs\mu\mu}$ & $+0.14 \pm 0.23$ \\
$C_9^{\prime bs\mu\mu}$ & $-0.10 \pm 0.34$ \\
$C_{10}^{\prime bs\mu\mu}$ & $-0.33 \pm 0.23$ \\
$C_9^{bsee}$ & $-0.24 \pm 1.17$ \\
$C_{10}^{bsee}$ & $-0.24 \pm 0.78$ \\
\end{tabular}\qquad
\begin{tabular}{l|rrrrrr}
& $C_9^{bs\mu\mu}$ & $C_{10}^{bs\mu\mu}$ & $C_9^{\prime bs\mu\mu}$ & $C_{10}^{\prime bs\mu\mu}$ & $C_9^{bsee}$ & $C_{10}^{bsee}$ \\
\hline
$C_9^{bs\mu\mu}$ & $1\phantom{.00}$ & $0.27$ & $0.22$ & $0.36$ & $-0.07$ & $-0.17$ \\
$C_{10}^{bs\mu\mu}$ &  & $1\phantom{.00}$ & $0.38$ & $0.68$ & $-0.33$ & $-0.01$ \\
$C_9^{\prime bs\mu\mu}$ &  &  & $1\phantom{.00}$ & $0.70$ & $0.17$ & $0.21$ \\
$C_{10}^{\prime bs\mu\mu}$ &  &  &  & $1\phantom{.00}$ & $-0.32$ & $-0.13$ \\
$C_9^{bsee}$ &  &  &  &  & $1\phantom{.00}$ & $0.90$ \\
$C_{10}^{bsee}$ &  &  &  &  &  & $1\phantom{.00}$ \\
\end{tabular}
\caption{Best fit values, uncertainties, and correlation matrix of the six-parameter fit to the Wilson coefficients $C_9^{bs\mu\mu}$, $C_{10}^{bs\mu\mu}$, $C_9^{\prime bs\mu\mu}$, $C_{10}^{\prime bs\mu\mu}$, $C_9^{bsee}$, and $C_{10}^{bsee}$ including all observables.}
\label{tab:6parameter}
\end{table}

We find similar results in the six parameter scenario. The parameters of the multivariate Gaussian that approximates the likelihood function in the vicinity of the best fit point of the global fit is reported in Table~\ref{tab:6parameter}. The results for the muon specific Wilson coefficients are very similar to the four prameter fit discussed above. New physics effects in the electron-specific Wilson coefficients $C_9^{bsee}$ and $C_{10}^{bsee}$ are complatible with zero. The uncertainties of $C_9^{bsee}$ and $C_{10}^{bsee}$ are large and highly correlated.

\subsection{Complex Wilson Coefficients}

\afterpage{%
\begin{figure}[tb]
\centering
\includegraphics[width=0.48\textwidth]{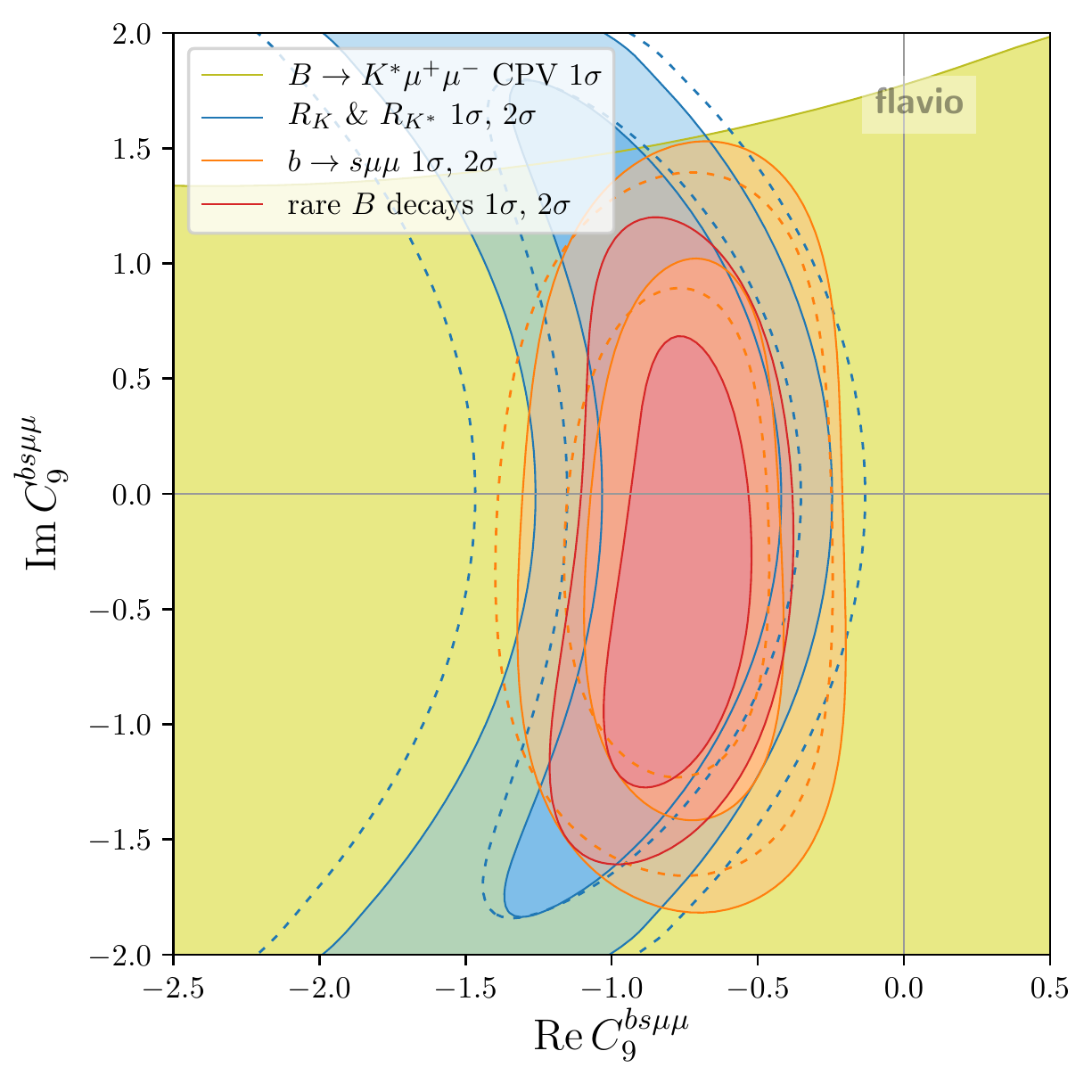} ~~~
\includegraphics[width=0.48\textwidth]{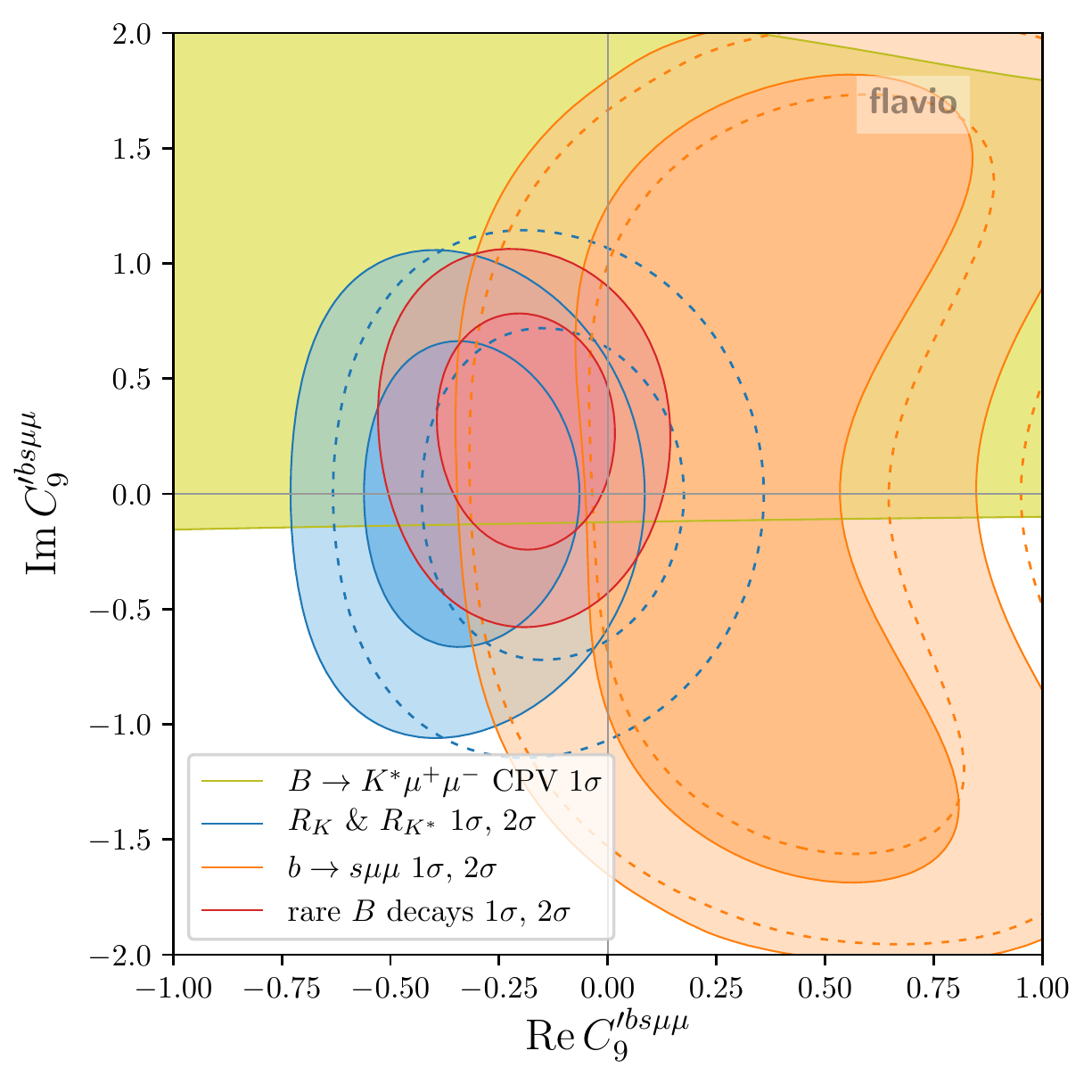} \\[12pt]
\includegraphics[width=0.48\textwidth]{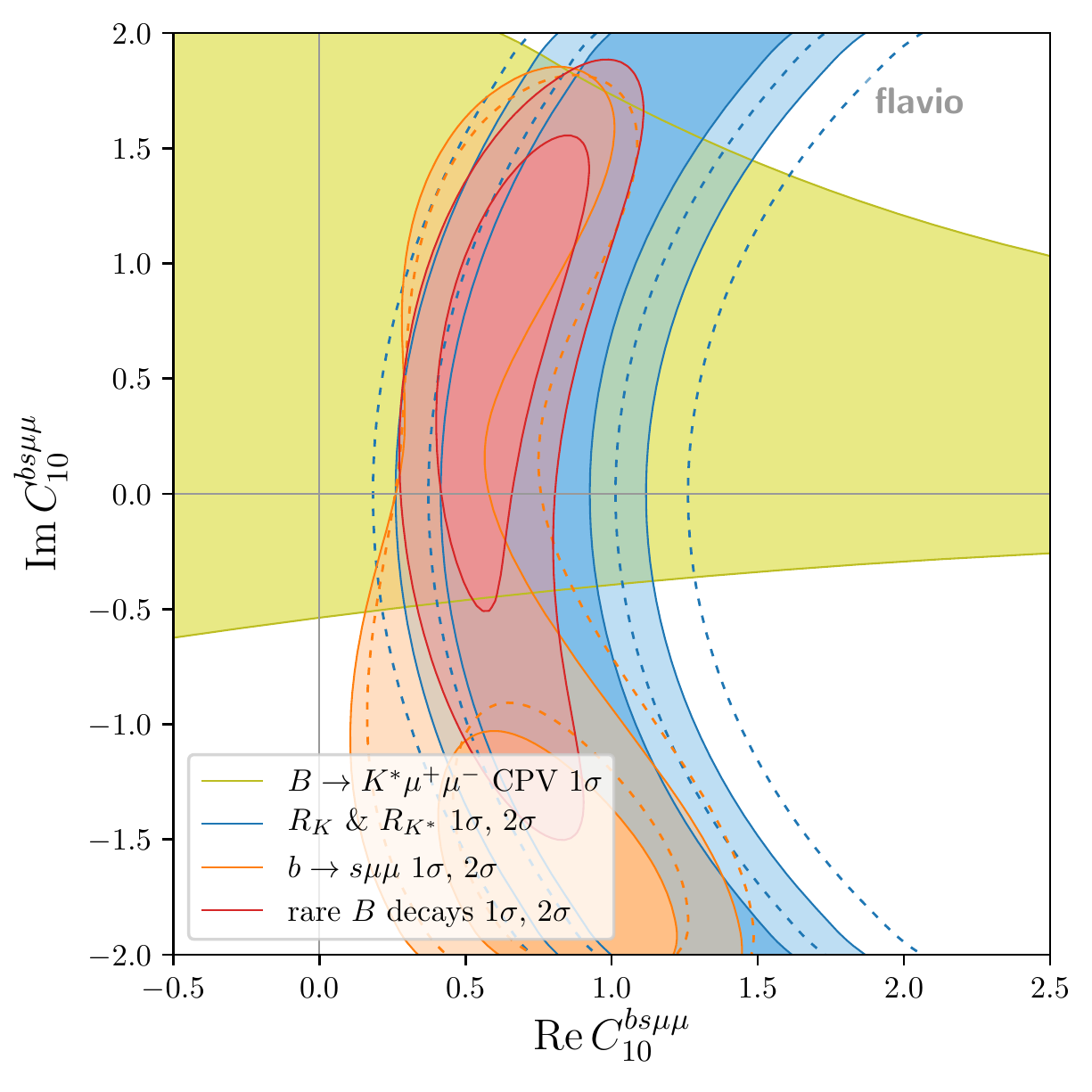} ~~~
\includegraphics[width=0.48\textwidth]{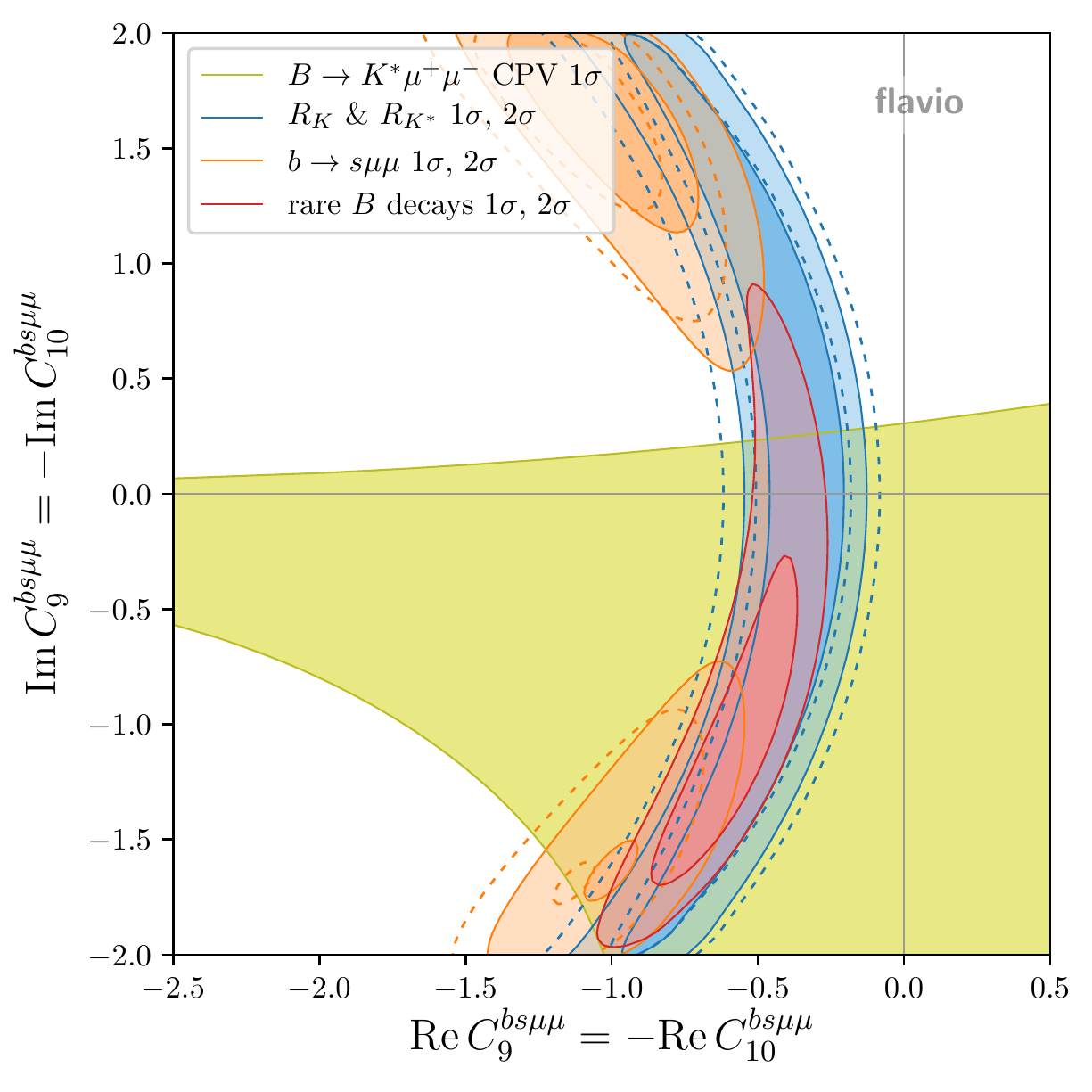}
\caption{Constraints in the planes of complex $C_9^{bs\mu\mu}$ (top left), $C_9^{\prime\,bs\mu\mu}$ (top right), $C_{10}^{bs\mu\mu}$ (bottom left), and $C_9^{bs\mu\mu} = - C_{10}^{bs\mu\mu}$ (bottom right). Shown separately are the constraints from LFU observables, CP conserving $b \to s \mu\mu$ observables, the $B \to K^* \mu^+\mu^-$ CP asymmetries, and the global fit. The dashed lines show the constraints before the recent updates~\cite{Aaij:2021vac,Aaij:2021pkz}.}
\label{fig:complex_WCs}
\end{figure}
\clearpage}

In the presence of new physics, the contributions to the flavor changing Wilson coefficients can generically be CP violating.
While the observables that show tensions with SM predictions are CP conserving, it is interesting to investigate the impact that imaginary parts of Wilson coefficients have on the fit, and to which extent imaginary parts are constrained by existing data
(see also~\cite{Carvunis:2021jga} for a recent study that considers complex Wilson coefficients).

In Figure~\ref{fig:complex_WCs} we show constraints in the planes of complex $C_9^{bs\mu\mu}$ (top left), $C_9^{\prime\,bs\mu\mu}$ (top right), $C_{10}^{bs\mu\mu}$ (bottom left), and $C_9^{bs\mu\mu} = - C_{10}^{bs\mu\mu}$ (bottom right). Shown separately are the constraints from LFU observables, CP conserving $b \to s \mu\mu$ observables, the $B \to K^* \mu^+\mu^-$ CP asymmetries from~\cite{Aaij:2015oid}, and the global fit.

In the case of $C_9^{bs\mu\mu}$, the experimental data does not lead to relevant constraints on the imaginary part of the Wilson coefficient, yet. In fact the strongest constraint on Im$(C_9^{bs\mu\mu})$ arises due to the fact that a sizeable imaginary part universally enhances the $b \to s \mu\mu$ rates. We observe that the other scenarios Im$(C_9^{\prime\,bs\mu\mu})$, Im$(C_{10}^{bs\mu\mu})$, and Im$(C_{9}^{bs\mu\mu}) = $Im$(C_{10}^{bs\mu\mu})$ are already being constrained by the experimental data on the CP asymmetries. Still, the current measurements do leave room for imaginary parts that are at least as large as the corresponding real parts. All imaginary parts are compatible with zero at the $2\sigma$ level.
The best fit points of the real part of the Wilson coefficients are very close to the values that we obtain setting the imaginary parts to zero.

\section{Predictions for LFU Observables and CP Asymmetries} \label{sec:predictions}

As discussed in the previous section, several new physics Wilson coefficients (or combinations of Wilson coefficients) can significantly improve the agreement between data and theory predictions. The various best fit points show comparable pulls, and it is therefore interesting to identify predictions that allow us to distinguish the new physics scenarios.

We consider six different two parameter new physics scenarios: (i) $\text{Re}\,C^{bs\mu\mu}_9$ \& $\text{Im}\,C^{bs\mu\mu}_9$, (ii) $\text{Re}\,C^{bs\mu\mu}_{10}$ \& $\text{Im}\,C^{bs\mu\mu}_{10}$, (iii) $\text{Re}\,C^{bs\mu\mu}_9=-\text{Re}\,C^{bs\mu\mu}_{10}$ \& $\text{Im}\,C^{bs\mu\mu}_9=-\text{Im}\,C^{bs\mu\mu}_{10}$, (iv) $C^{bs\mu\mu}_9$ \& $C^{bs\mu\mu}_{10}$, (v) $C^\text{univ.}_9$ \& $\Delta C^{bs\mu\mu}_9 = -C^{bs\mu\mu}_{10}$, and (vi) $C^{bs\mu\mu}_9$ \& $C^{\prime\,bs\mu\mu}_9$.
In each of these cases, we sample the likelihood of the Wilson coefficients and show in Table~\ref{tab:predictions} the predictions for several observables.

\begin{table}[tbp]
\centering
\renewcommand{\arraystretch}{1.5}
\rowcolors{2}{gray!15}{white}
\addtolength{\tabcolsep}{-0.11pt} 
\begin{tabularx}{\textwidth}{ccccccc}
\toprule
\rowcolor{white}
  & (i)  & (ii)  & (iii)  & (iv)  & (v)  & (vi) \\
\midrule
$R_K^{[1.1,6.0]}$ & $+0.85_{-0.03}^{+0.03}$ & $+0.87_{-0.03}^{+0.03}$ & $+0.83_{-0.04}^{+0.03}$ & $+0.83_{-0.04}^{+0.04}$ & $+0.82_{-0.04}^{+0.04}$ & $+0.86_{-0.04}^{+0.04}$ \\
$R_K^{[14.18,19.0]}$ & $+0.85_{-0.03}^{+0.03}$ & $+0.88_{-0.03}^{+0.03}$ & $+0.83_{-0.04}^{+0.03}$ & $+0.83_{-0.04}^{+0.04}$ & $+0.82_{-0.04}^{+0.04}$ & $+0.86_{-0.04}^{+0.04}$ \\
$R_{K^\ast}^{[0.045,1.1]}$ & $+0.90_{-0.01}^{+0.01}$ & $+0.88_{-0.01}^{+0.01}$ & $+0.89_{-0.01}^{+0.01}$ & $+0.89_{-0.02}^{+0.01}$ & $+0.88_{-0.01}^{+0.01}$ & $+0.89_{-0.01}^{+0.02}$ \\
$R_{K^\ast}^{[1.1,6.0]}$ & $+0.89_{-0.02}^{+0.03}$ & $+0.85_{-0.03}^{+0.03}$ & $+0.84_{-0.04}^{+0.04}$ & $+0.85_{-0.04}^{+0.04}$ & $+0.82_{-0.03}^{+0.04}$ & $+0.83_{-0.04}^{+0.05}$ \\
$R_{K^\ast}^{[15,19]}$ & $+0.85_{-0.03}^{+0.03}$ & $+0.86_{-0.03}^{+0.03}$ & $+0.82_{-0.04}^{+0.03}$ & $+0.82_{-0.04}^{+0.04}$ & $+0.81_{-0.04}^{+0.04}$ & $+0.79_{-0.04}^{+0.05}$ \\
$R_\phi^{[1.0,6.0]}$ & $+0.88_{-0.02}^{+0.03}$ & $+0.85_{-0.03}^{+0.03}$ & $+0.84_{-0.04}^{+0.05}$ & $+0.84_{-0.04}^{+0.04}$ & $+0.82_{-0.04}^{+0.04}$ & $+0.83_{-0.04}^{+0.05}$ \\
$R_\phi^{[15,19]}$ & $+0.85_{-0.03}^{+0.04}$ & $+0.87_{-0.03}^{+0.03}$ & $+0.83_{-0.04}^{+0.03}$ & $+0.82_{-0.04}^{+0.04}$ & $+0.81_{-0.04}^{+0.04}$ & $+0.79_{-0.04}^{+0.05}$ \\
\midrule
$D_{P_5^\prime}^{[1.0,6.0]}$ & $+0.19_{-0.04}^{+0.07}$ & $-0.02_{-0.01}^{+0.02}$ & $+0.06_{-0.03}^{+0.02}$ & $+0.13_{-0.07}^{+0.04}$ & $+0.09_{-0.02}^{+0.03}$ & $+0.21_{-0.05}^{+0.07}$ \\
$D_{P_4^\prime}^{[1.0,6.0]}$ & $+0.01_{-0.01}^{+0.00}$ & $+0.03_{-0.01}^{+0.01}$ & $+0.03_{-0.01}^{+0.01}$ & $+0.02_{-0.01}^{+0.01}$ & $+0.03_{-0.01}^{+0.01}$ & $+0.02_{-0.01}^{+0.01}$ \\
$D_{A_\text{FB}}^{[1.0,6.0]}$ & $-0.05_{-0.02}^{+0.01}$ & $+0.00_{-0.00}^{+0.00}$ & $-0.02_{-0.01}^{+0.02}$ & $-0.04_{-0.01}^{+0.02}$ & $-0.03_{-0.01}^{+0.01}$ & $-0.06_{-0.02}^{+0.02}$ \\
\midrule
$A_{7}^{[1.1,6]}$ & $+0.00_{-0.00}^{+0.00}$ & $-0.06_{-0.03}^{+0.07}$ & $-0.09_{-0.03}^{+0.05}$ & $+0.00_{-0.00}^{+0.00}$ & $+0.00_{-0.00}^{+0.00}$ & $+0.00_{-0.00}^{+0.00}$ \\
$A_{8}^{[1.1,6]}$ & $-0.02_{-0.02}^{+0.03}$ & $+0.00_{-0.00}^{+0.00}$ & $-0.05_{-0.02}^{+0.03}$ & $+0.00_{-0.00}^{+0.00}$ & $+0.00_{-0.00}^{+0.00}$ & $+0.00_{-0.00}^{+0.00}$ \\

\bottomrule
\end{tabularx}
\addtolength{\tabcolsep}{-4pt} 
\caption{Predictions for lepton flavor universality observables and CP asymmetries in global fits of 2D new-physics scenarios as shown in figures~\ref{fig:2d1}, \ref{fig:2d2}, and \ref{fig:complex_WCs}:
(i) $\text{Re}\,C^{bs\mu\mu}_9$ \& $\text{Im}\,C^{bs\mu\mu}_9$,
(ii) $\text{Re}\,C^{bs\mu\mu}_{10}$ \& $\text{Im}\,C^{bs\mu\mu}_{10}$,
(iii) $\text{Re}\,C^{bs\mu\mu}_9=-\text{Re}\,C^{bs\mu\mu}_{10}$ \& $\text{Im}\,C^{bs\mu\mu}_9=-\text{Im}\,C^{bs\mu\mu}_{10}$,
(iv) $C^{bs\mu\mu}_9$ \& $C^{bs\mu\mu}_{10}$,
(v) $C^\text{univ.}_9$ \& $\Delta C^{bs\mu\mu}_9 = -C^{bs\mu\mu}_{10}$,
(vi) $C^{bs\mu\mu}_9$ \& $C^{\prime\,bs\mu\mu}_9$.
The superscripts on the observables indicate the $q^2$ range in GeV$^2$.
}
\label{tab:predictions}
\end{table}

The first set of rows shows the predictions for the LFU ratios $R_K$, $R_{K^*}$, and $R_{\phi}$ both at low $q^2$ and at high $q^2$. Overall, the predictions are fairly similar in all the considered new physics scenarios.
Given the precise measurement of $R_K$ at low $q^2$ that enters the global fits, all scenarios reproduce the measurement of $\simeq 0.85$ at the $1\sigma$ level. The predicted values for all other LFU ratios are similar in all scenarios (i) - (vi). The central values are all expected between $0.8$ and $0.9$. This is in particularly true for $R_{K^*}$ where the current experimental result is considerably lower.

The second set of rows shows predictions for LFU differences of $B \to K^* \mu^+\mu^-$ angular observables: $D_{P_5^\prime}$, $D_{P_4^\prime}$, and $D_{A_\text{FB}}$. Here we find significant differences in the various scenarios. In particular, precise measurements of $D_{P_5^\prime}$ will allow to narrow down new physics scenarios.

The last set of rows shows predictions for the $B \to K^* \mu^+\mu^-$ CP asymmetries $A_7$ and $A_8$.
The CP asymmetries remain close to zero (i.e. SM-like) in the scenarios (iv)-(vi) as they do not contain any new sources of CP violation. In scenarios (i)-(iii), $A_7$ and $A_8$ can be non-zero. Interestingly, an imaginary part of $C^{bs\mu\mu}_{9}$ leads to an effect in $A_8$, while an imaginary part of $C^{bs\mu\mu}_{10}$ leads to an effect in $A_7$. The predicted ranges for $A_7$ and $A_8$ can already be probed with run 2 data.

\begin{figure}[tb]
\centering
\includegraphics[width=0.8\textwidth]{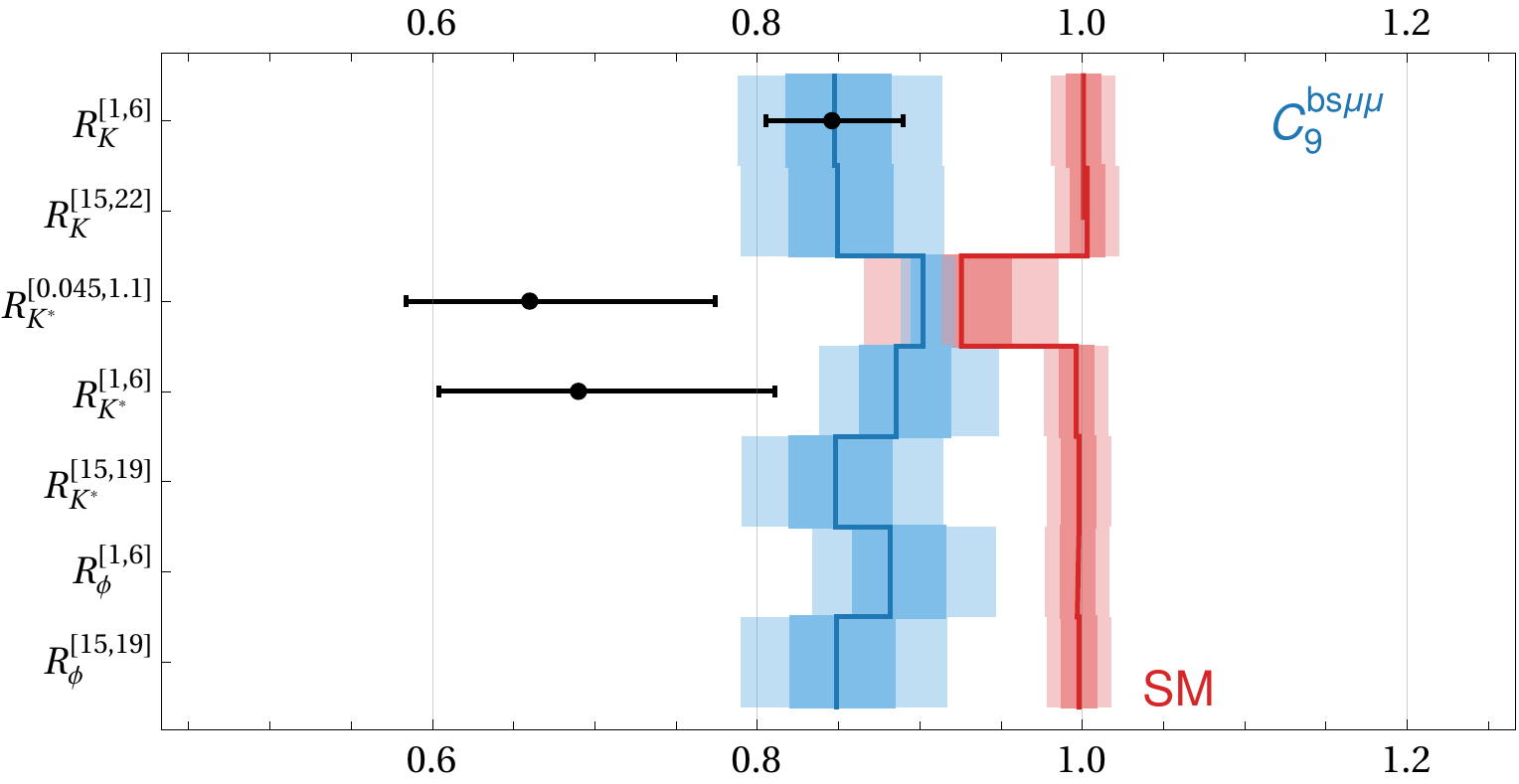} \\[12pt]
\includegraphics[width=0.8\textwidth]{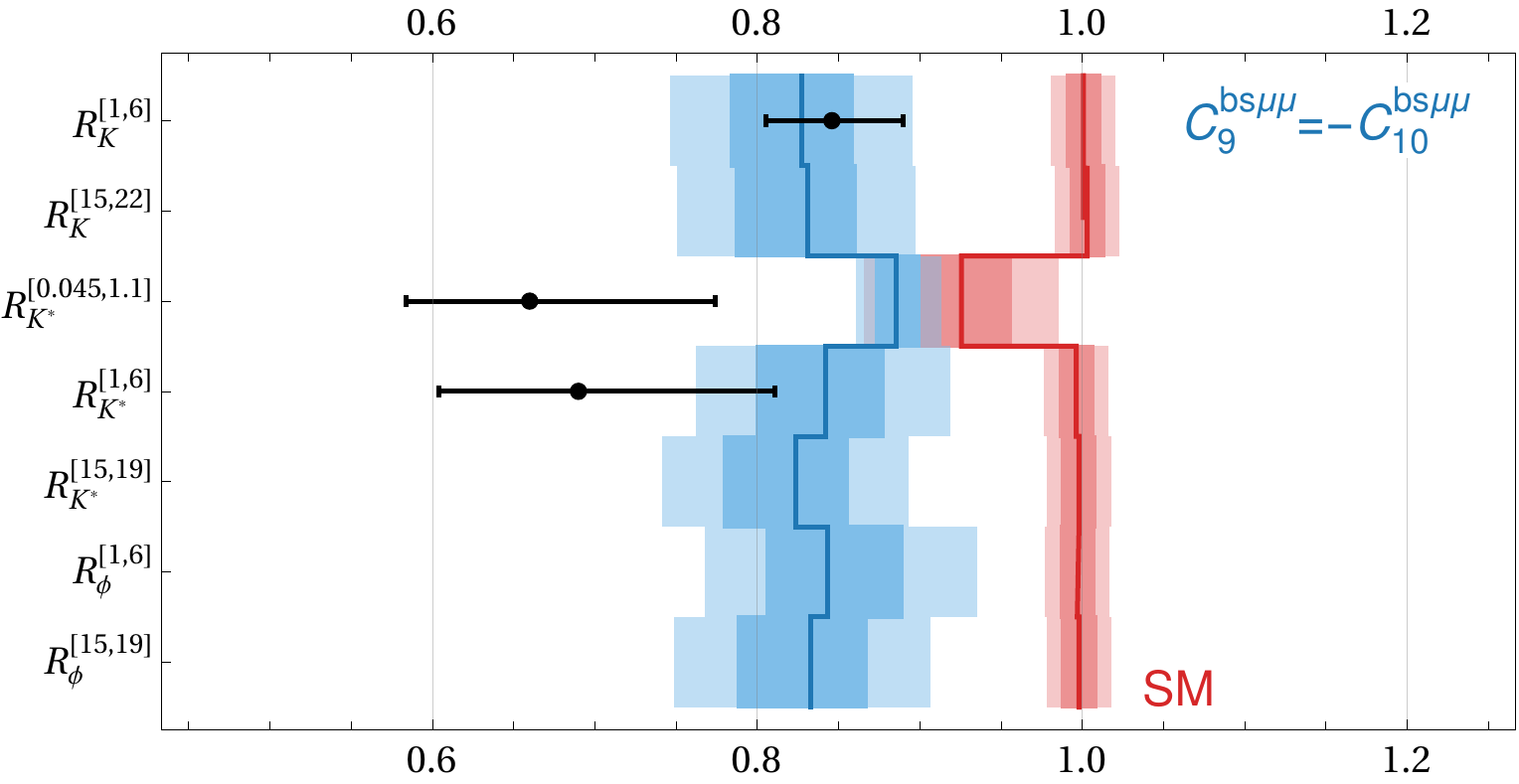} \\[12pt]
\includegraphics[width=0.8\textwidth]{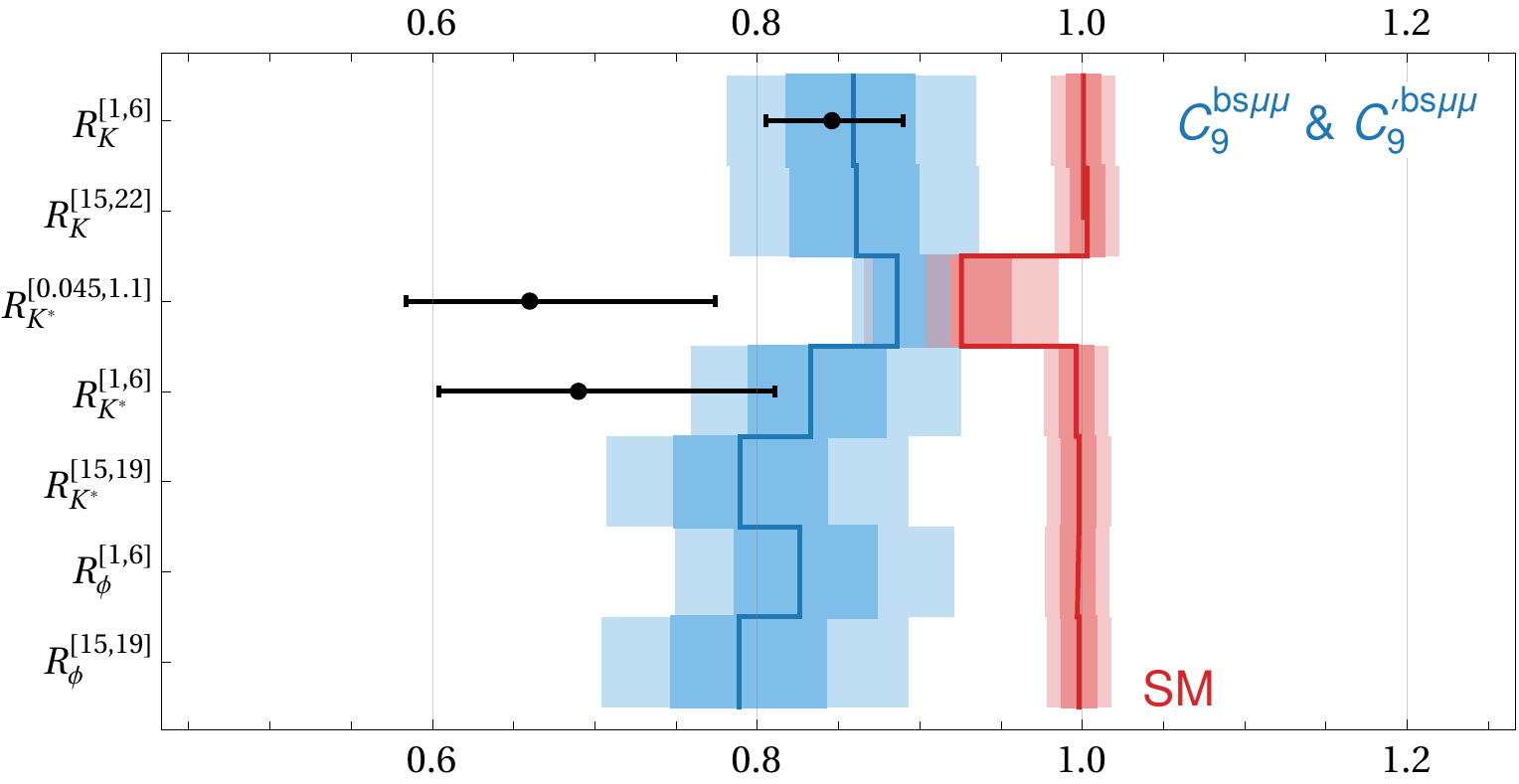}
\caption{Predictions for the LFU ratios $R_K$, $R_{K^*}$, and $R_{\phi}$ in three new physics scenarios and the SM. For comparison the current measurements from LHCb~\cite{Aaij:2019wad,Aaij:2017vbb} are shown as well.}
\label{fig:predictions_1}
\end{figure}
\begin{figure}[tb]
\centering
\includegraphics[width=0.8\textwidth]{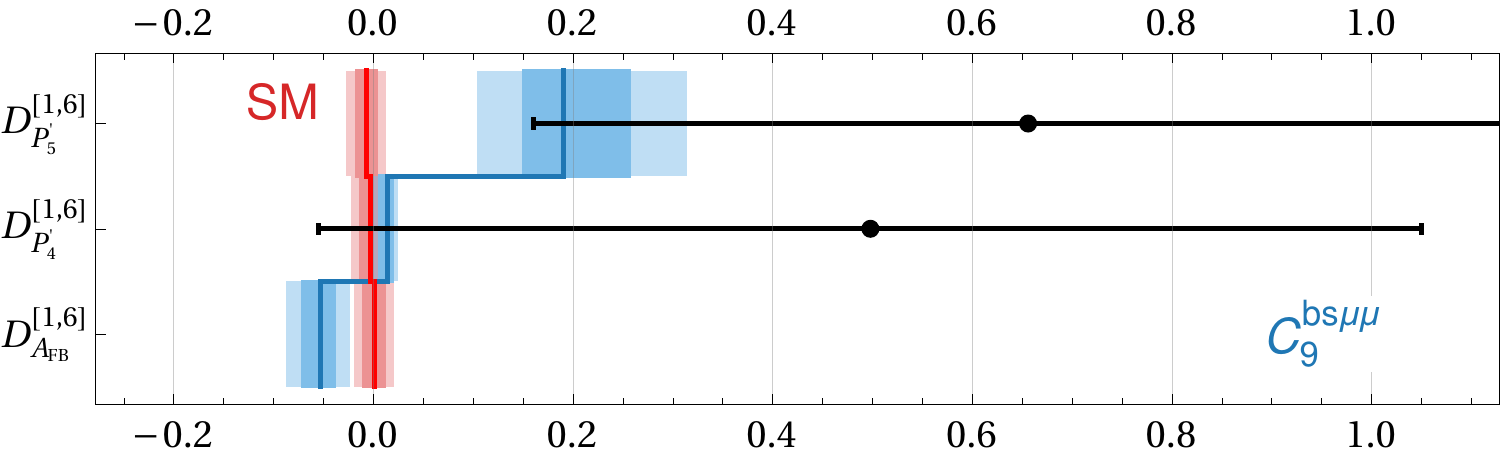} \\[12pt]
\includegraphics[width=0.8\textwidth]{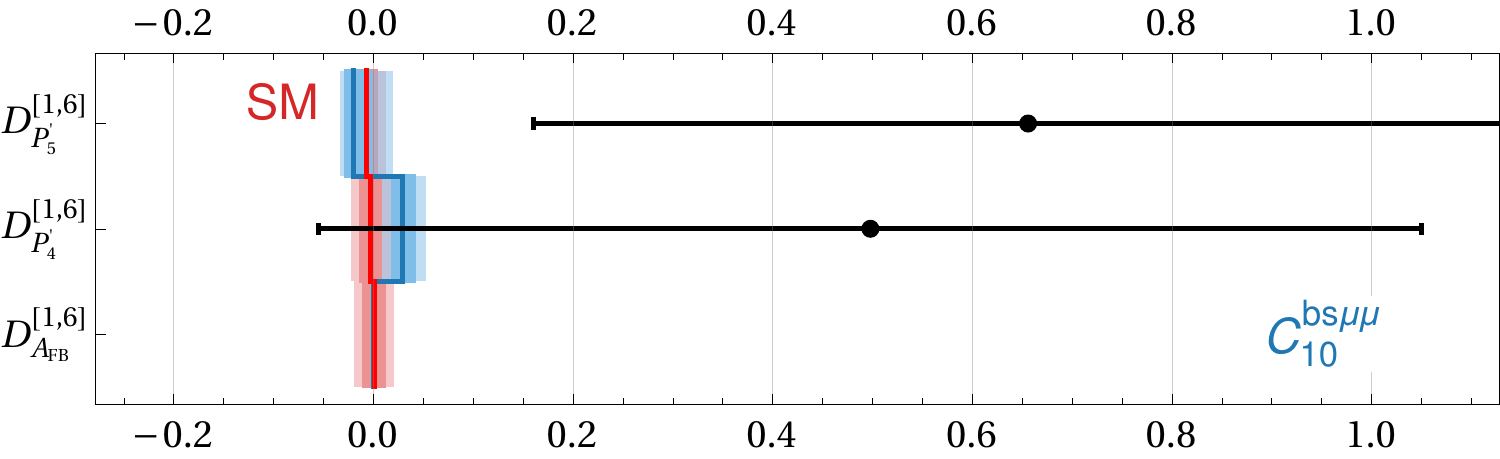} \\[12pt]
\includegraphics[width=0.8\textwidth]{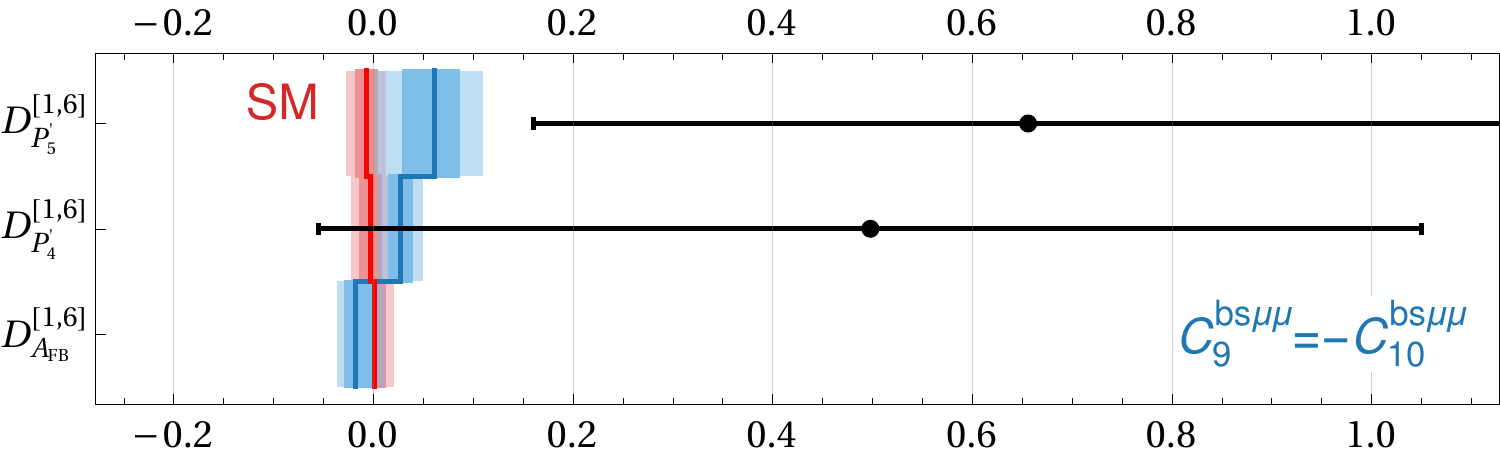}
\caption{Predictions for the LFU differences $D_{P_5^\prime}$, $D_{P_4^\prime}$, and $D_{A_\text{FB}}$ in three new physics scenarios and the SM. For comparison the current measurements from Belle~\cite{Wehle:2016yoi} are shown as well.}
\label{fig:predictions_2}
\end{figure}
\clearpage

\begin{figure}[tb]
\centering
\includegraphics[width=0.8\textwidth]{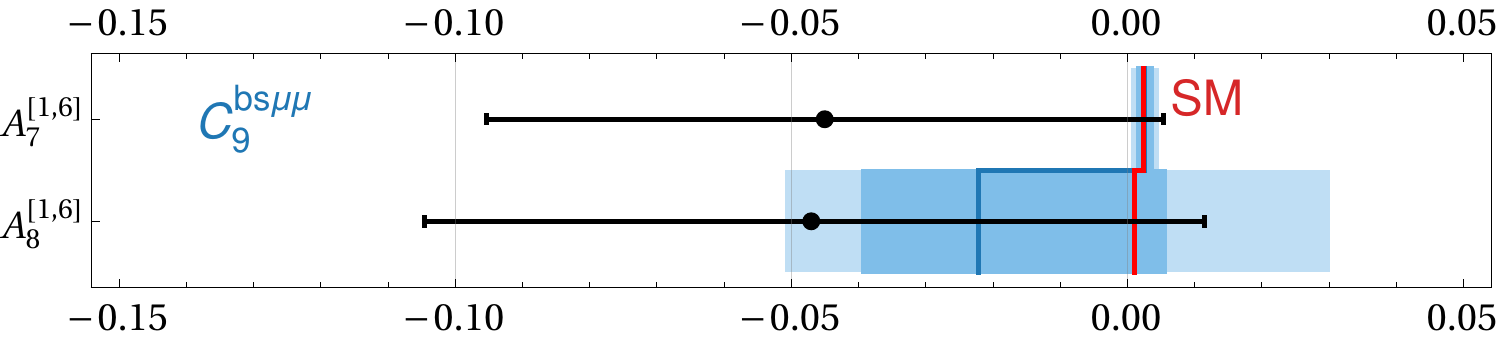} \\[12pt]
\includegraphics[width=0.8\textwidth]{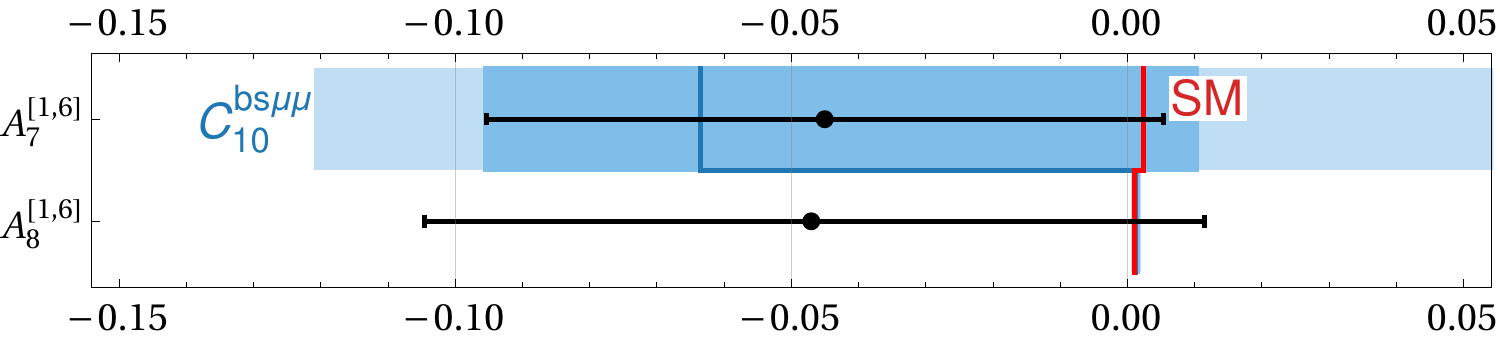} \\[12pt]
\includegraphics[width=0.8\textwidth]{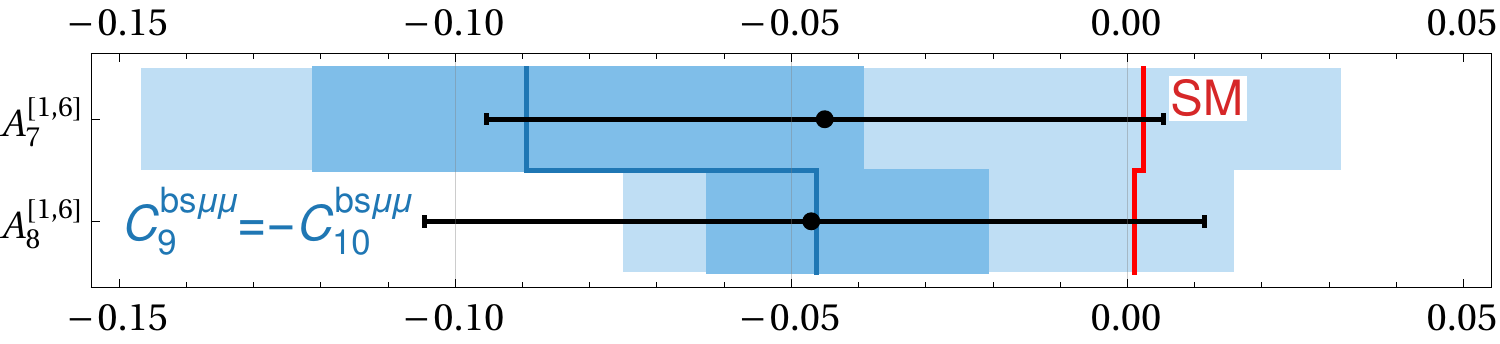}
\caption{Predictions for the CP asymmetries $A_7$ and $A_8$ in three new physics scenarios and the SM. For comparison the current measurements from LHCb~\cite{Aaij:2015oid} are shown as well.}
\label{fig:predictions_3}
\end{figure}

In Figures~\ref{fig:predictions_1},~\ref{fig:predictions_2}, and~\ref{fig:predictions_3}, we show the most distinctive cases in graphical form. The plots of Figure~\ref{fig:predictions_1} contain the predictions for the LFU ratios in scenarios (i), (iii), and (iv). The new physics predictions are compared to the SM predictions (with uncertainties from~\cite{Bordone:2016gaq}) and the current experimental results~\cite{Aaij:2017vbb,Aaij:2021vac}. Similarly, the plots of Figure~\ref{fig:predictions_2} show predictions and experimental results~\cite{Wehle:2016yoi} for the LFU differences in scenarios (i), (ii), and (iii). The uncertainties of the SM predictions are illustrated with $\pm 0.01$. Finally, the plots of Figure~\ref{fig:predictions_3} show the CP Asymmetries in the scenarios with imaginary parts (i), (ii), and (iii). The tiny SM uncertainties are neglected and the experimental results are taken from~\cite{Aaij:2015oid}.
The plots clearly show the discrimination power of the different observables.

\section{Conclusions} \label{sec:conclusions}

With the recent updates of $R_K$ and BR$(B_s \to \mu^+\mu^-)$ by LHCb, the case for new physics in rare B decays has been further strengthened. Our improved global fit shows very strong preference for the muon specific Wilson coefficients $C_9^{bs\mu\mu} \simeq -0.73$ or $C_9^{bs\mu\mu} = -C_{10}^{bs\mu\mu} \simeq -0.39$. Even if only the theoretically clean LFU observables and BR$(B_s \to \mu^+\mu^-)$ are considered, muon specific $C_{10}^{bs\mu\mu} \simeq 0.60$ or $C_9^{bs\mu\mu}=-C_{10}^{bs\mu\mu} \simeq -0.35$ improve over the Standard Model by $\sqrt{\Delta \chi^2} \simeq 4.7\sigma$ and $\sqrt{\Delta \chi^2} \simeq 4.6\sigma$, respectively.
We have also investigated complex Wilson coefficients and find relevant constraints on the imaginary parts of $C_{10}^{bs\mu\mu}$ and $C_{9}^{\prime\,bs\mu\mu}$ from the experimental results on the $B \to K^* \mu^+\mu^-$ CP asymmetries.

Finally, we give new physics predictions for a large set of observables including LFU ratios, LFU differences of CP averaged $B \to K^* \mu^+\mu^-$ observables, and $B \to K^* \mu^+\mu^-$ CP asymmetries. Future more precise measurements of these observables will allow us to distinguish between different new physics scenarios.

\paragraph{Note Added:} Another model independent interpretation of the new results can be found in~\cite{1853015}. First interpretations in new physics models have been presented in~\cite{1853087, 1853016}.

\section*{Acknowledgements}
The research of W. A. is supported by the U.S. Department of Energy grant number DE-SC0010107.
The work of P.~S.\ is supported by the Swiss National Science Foundation grant 200020175449/1.

\appendix
\section{Appendix: $B_q\to\mu^+\mu^-$ Combination}\label{app:Bsmumu}

We combine the ATLAS, CMS, and the recent LHCb measurement of the $B^0\to\mu^+\mu^-$ and $B_s\to\mu^+\mu^-$ branching ratios~\cite{Aaboud:2018mst, Sirunyan:2019xdu, LHCb:2021awg,LHCb:2021vsc}, following a procedure similar as in~\cite{Aebischer:2019mlg}.

\begin{figure}
\centering
\includegraphics[width=0.8\textwidth]{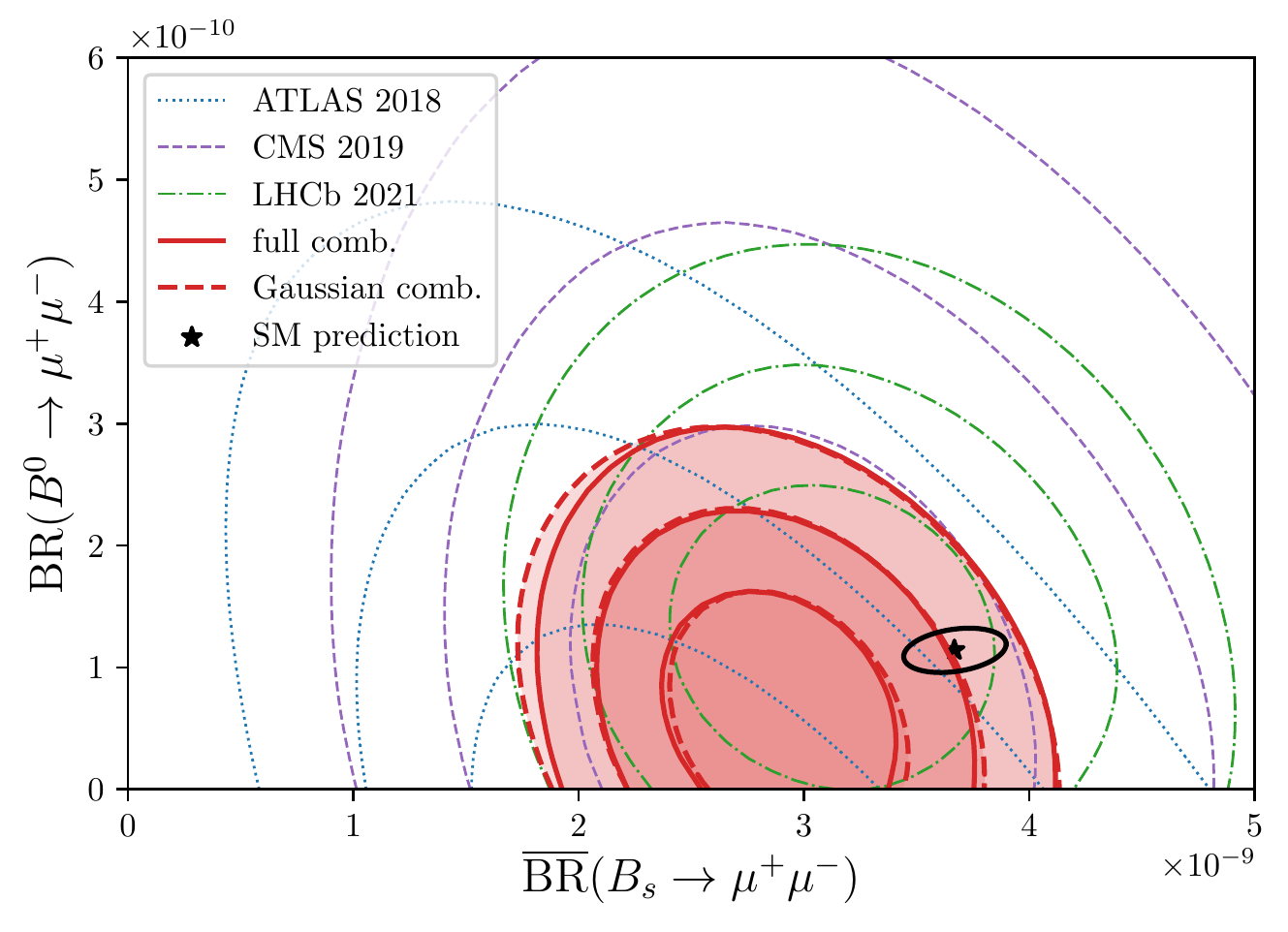}
\caption{Likelihood contours in the plane of BR$(B^0\to\mu^+\mu^-)$ and BR$(B_s\to\mu^+\mu^-)$ from the individual ATLAS, CMS, and LHCb measurements (thin contours), our combination (thick solid contours), and the Gaussian approximation (thick dashed contours). Also shown are the SM predictions and their 1$\sigma$ correlated uncertainties.}
\label{fig:Bsmumu}
\end{figure}

Since the $B^0$ and $B_s$ have a similar mass the measurements of the $B^0\to\mu^+\mu^-$ and $B_s\to\mu^+\mu^-$ branching ratios are correlated and the experimental results are given by two-dimensional likelihoods. We combine them assuming the likelihoods of different experiments are uncorrelated. The individual likelihoods are shown as thin lines in Fig.~\ref{fig:Bsmumu} while our combination is shown as thick solid red line. We also determine a Gaussian approximation (shown as thick dashed red line) and compare the experimental results to the SM predictions.

The two-dimensional Gaussian approximation is given by
\begin{align}
  \overline{\text{BR}}(B_s\to\mu^+\mu^-)_\text{exp}
  &= (2.93\pm0.35) \times 10^{-9},
 \\
 {\text{BR}}(B^0\to\mu^+\mu^-)_\text{exp}
 &= (0.56\pm0.70) \times 10^{-10},
\end{align}
with an error correlation of $\rho = -0.27$.

For the SM predictions, we use \verb|flavio| with default settings (The most relevant input parameters are the CKM elements $V_{cb} = (42.21 \pm 0.78)\times 10^{-3}$ and $V_{ub} = (3.73 \pm 0.14)\times 10^{-3}$ and the decay constants $f_{B_s} = (230.3\pm 1.3)$~MeV and $f_{B} = (190.0\pm 1.3)$~MeV~\cite{Aoki:2019cca})
\begin{align}
  \overline{\text{BR}}(B_s\to\mu^+\mu^-)_\text{SM}
  &= (3.67\pm0.15) \times 10^{-9},
 \\
 {\text{BR}}(B^0\to\mu^+\mu^-)_\text{SM}
 &= (1.14\pm0.12) \times 10^{-10},
\end{align}
with an error correlation of $\rho=+0.28$.

Comparing the SM predictions with the two dimensional experimental likelihood we get the following one-dimensional pulls\footnote{Here, the ``one-dimensional pull'' is $-2$ times the logarithm of the likelihood ratio at the SM vs.\ the experimental point, after the experimental uncertainties have been convoluted with the covariance of the SM uncertainties.}:
\begin{itemize}
  \item if both branching ratios are SM-like, $2.3\sigma$\footnote{Converting the likelihood ratio to a pull with two degrees of freedom, we get $1.8\sigma$.},
  \item if $B_s\to\mu^+\mu^-$ is SM-like and
  $B^0\to\mu^+\mu^-$ profiled over, $1.9\sigma$,
  \item if $B^0\to\mu^+\mu^-$ is SM-like and
  $B_s\to\mu^+\mu^-$ profiled over, $0.8\sigma$.
\end{itemize}

Given its prominent role in constraining new physics in $b \to s \mu\mu$ transitions, it is of great interest to have confidence regions for the $B_s\to\mu^+\mu^-$ branching ratio itself, fixing $B^0\to\mu^+\mu^-$ either to its SM central value or profiling over it. Using our two-dimensional likelihood, we find
\begin{align}
\overline{\text{BR}}(B_s\to\mu^+\mu^-)
&= (2.93^{+0.33}_{-0.35}) \times 10^{-9}
&& \text{BR}(B^0\to\mu^+\mu^-) \text{ profiled,}
 \\
 \overline{\text{BR}}(B_s\to\mu^+\mu^-)
&= (2.86^{+0.35}_{-0.32}) \times 10^{-9}
 && \text{BR}(B^0\to\mu^+\mu^-) \text{ SM-like.}
\end{align}
For $B^0\to\mu^+\mu^-$ we get analogously
\begin{align}
{\text{BR}}(B^0\to\mu^+\mu^-)
&= (0.56^{+0.70}_{-0.36}) \times 10^{-10}
&& \overline{\text{BR}}(B_s\to\mu^+\mu^-) \text{ profiled,}
 \\
 {\text{BR}}(B^0\to\mu^+\mu^-)
&= (0.24^{+0.72}_{-0.17}) \times 10^{-10}
 && \overline{\text{BR}}(B_s\to\mu^+\mu^-) \text{ SM-like.}
\end{align}

\section{Appendix: Details on Theory Uncertainties}

\subsection{Parameterization of Non-Factorizable Effects} \label{app:power}

We parameterize the non-factorizable effects in the decay amplitudes of semileptonic rare $B$ decays following~\cite{Altmannshofer:2014rta,Bharucha:2015bzk}.

For $B \to K$ decays, the Wilson coefficient $C_9^\text{eff}(q^2)$ is modified in the following way
\begin{equation}
\begin{array}{ll}
C_9^\text{eff}(q^2) \to C_9^\text{eff}(q^2) + a_K + b_K (q^2/\,\text{GeV}^2)
& \text{at low }q^2\,, \\

C_9^\text{eff}(q^2) \to C_9^\text{eff}(q^2) + c_K
& \text{at high }q^2\,,
\end{array}
\end{equation}
where low $q^2$ and high $q^2$ refers to di-lepton invariant masses below and above the narrow charmonium resonances, respectively. The central values of the complex parameters $a_K$, $b_K$, and $c_K$ are set to zero and the $1\sigma$ uncertainties enclose the effects considered in~\cite{Khodjamirian:2010vf,Beylich:2011aq,Khodjamirian:2012rm}
\begin{align}
 \text{Re}(a_K) &= 0.0 \pm 0.08 ~,& \text{Re}(b_K) &= 0.0 \pm 0.03 ~,& \text{Re}(c_K) &= 0.0 \pm 0.2~, \\
 \text{Im}(a_K) &= 0.0 \pm 0.08 ~,& \text{Im}(b_K) &= 0.0 \pm 0.03 ~,& \text{Im}(c_K) &= 0.0 \pm 0.2~.
\end{align}
We use the same ranges for $B^+ \to K^+$ and $B^0 \to K^0$ decays and assume that the corresponding coefficients are correlated by $+99\%$ due to iso-spin symmetry.

For $B \to K^*$ and $B_s \to \phi$ decays we use the following parameterization
\begin{equation}
\begin{array}{rl}
C_7^\text{eff}(q^2) &\to C_7^\text{eff}(q^2) + a_{0,-} + b_{0,-} (q^2/\,\text{GeV}^2)\\
C_7^\prime &\to C_7^\prime + a_+ + b_+ (q^2/\,\text{GeV}^2)
\end{array} \qquad \text{at low }q^2\,,
\end{equation}
where the replacement of $C_7^\text{eff}$ is performed only in the $\lambda = 0,-$ helicity amplitudes, and the replacement of $C_7^\prime$ only in the $\lambda = +$ amplitude. Furthermore, we have
\begin{equation}
C_9^\text{eff}(q^2) \to C_9^\text{eff}(q^2) + c_\lambda \qquad \text{at high }q^2\,,
\end{equation}
in all the helicity amplitudes. We use the following values for the hadronic parameters
\begin{align}
 \text{Re}(a_+) &= 0.0 \pm 0.004 ~,& \text{Re}(b_+) &= 0.0 \pm 0.005 ~,& \text{Re}(c_+) &= 0.0 \pm 0.3 ~, \\
 \text{Im}(a_+) &= 0.0 \pm 0.004 ~,& \text{Im}(b_+) &= 0.0 \pm 0.005 ~,& \text{Im}(c_+) &= 0.0 \pm 0.3 ~, \\
 \text{Re}(a_-) &= 0.0 \pm 0.015 ~,& \text{Re}(b_-) &= 0.0 \pm 0.01 ~,& \text{Re}(c_-) &= 0.0 \pm 0.3 ~, \\
 \text{Im}(a_-) &= 0.0 \pm 0.015 ~,& \text{Im}(b_-) &= 0.0 \pm 0.01 ~,& \text{Im}(c_-) &= 0.0 \pm 0.3 ~, \\
 \text{Re}(a_0) &= 0.0 \pm 0.12 ~,& \text{Re}(b_0) &= 0.0 \pm 0.05 ~,& \text{Re}(c_0) &= 0.0 \pm 0.3 ~, \\
 \text{Im}(a_0) &= 0.0 \pm 0.12 ~,& \text{Im}(b_0) &= 0.0 \pm 0.05 ~,& \text{Im}(c_0) &= 0.0 \pm 0.3 ~.
\end{align}
The same ranges of the parameters are considered for $B^0 \to K^{*\,0}$, $B^+ \to K^{*\,+}$, and $B_s \to \phi$ decays. A $+99\%$ correlation is assumed between the $B^0 \to K^{*\,0}$ and $B^+ \to K^{*\,+}$ coefficients (due to iso-spin), and a $+90\%$ correlation between the coefficients for the $B_s \to \phi$ decay and the $B \to K^*$ decays (due to $SU(3)$ symmetry).

The above treatment of the non-factorizable effects is implemented in \verb|flavio| since version \verb|1.0|.

\subsection{Implementation of the New Physics Dependence} \label{app:uncertainties}

The decay amplitudes of rare semileptonic $b$ hadron decays are linear functions of the Wilson coefficients. Thus, in the presence of new physics, the angular coefficients in the differential decay rates are second order polynomials in the new physics Wilson coefficients. Any observable $O_k$ in rare semileptonic decays that we consider can therefore be written as a function of second order polynomials $p_i$
\begin{equation}
 O_k = f_k(p_1, p_2, ..., p_n)\,.
\end{equation}
For example, binned branching ratios are given directly in terms of a single second order polynomial, $f_k(p_1) = p_1$. The CP averaged angular observables $S_i$, the CP asymmetries $A_i$, and the LFU ratios are ratios of two second order polynomials $f_k(p_1,p_2) = p_1/p_2$. The angular observable $P_5^\prime$ has the form $f_k(p_1,p_2) = p_1/\sqrt{p_2(1-p_2)}$, and so on. The polynomials can be written in terms of a vector product
\begin{equation} \label{eq:polynomial}
 p_i = \vec{p}_i \cdot \vec{V} = a_i +  \epsilon (\vec{b}_i \cdot \vec{C}) + \epsilon^2 (\vec{c}_i\cdot \vec{D})\,,
\end{equation}
where $\vec{C}=(C_1,C_2,...,C_M)^T$ is a vector of new physics Wilson coefficients and $\vec{D}=\text{vec}(\vec{C}\otimes\vec{C})$ is a vector of products of Wilson coefficients.\footnote{For a $n$ component vector $\vec v = (v_1,...,v_n)^T$ and a $m$ component vector $\vec u = (u_1,...,u_m)^T$ we define the $n\times m$ component vector $$\text{vec}(\vec{v}\otimes\vec{u}) = (v_1u_1,v_1u_2,...,v_1u_m,v_2u_1,v_2u_2,...,v_2u_m,...,v_nu_1,v_nu_2,...,v_nu_m)^T~.$$} The vector $\vec{V} = (1, \epsilon \vec{C}^T, \epsilon^2 \vec{D}^T)^T$ is independent of the considered observable and contains the information about the new physics. The factors of $\epsilon$ are introduced to track the order in the Wilson coefficients and they will be set to $\epsilon=1$ in the end.

The vectors $\vec{p}_i = (a_i , \vec b_i^T, \vec c_i^T)^T$ in~\eqref{eq:polynomial} are independent of the new physics. They depend on the considered observable and are given in terms of known input parameters. For any set of observables we can determine the covariance matrix $\Sigma_{\vec p}$ for the corresponding set of vectors $\vec{p}_i$. If $N$ polynomials and $M$ Wilson coefficients are involved, $\Sigma_{\vec p}$ is a $N (1+M+M^2) \times N (1+M+M^2)$ matrix.\footnote{In practice, the size of the covariance matrix $\Sigma_{\vec p}$ can be slightly reduced by using the fact that only $M(M+1)/2$ out of the $M^2$ entries in $\vec{D}=\text{vec}(\vec{C}\otimes\vec{C})$ are independent and that usually some of the components of the $\vec p_i$ are exactly zero.} We infer $\Sigma_{\vec p}$ by varying the input parameters within uncertainties, assuming Gaussian distributions.

For branching ratios, the functions $f_k$ are the identity, the observables depend linearly on the $\vec{p}_i$, and the number of polynomials, $N$, is equal to the number of observables. In this case, the $N\times N$ theory covariance matrix $\Sigma_\text{th}$ that enters the $\chi^2$ function~\eqref{eq:chi2} can simply be written as (see e.g.~\cite{tong1990the})
\begin{equation} \label{eq:Sigma_th}
 \Sigma_\text{th} = (\mathbb{1}_{N}\otimes \vec V^T) \cdot \Sigma_{\vec p} \cdot (\vec V \otimes \mathbb{1}_{N}) \Big|_{\epsilon = 1} ~.
\end{equation}
This $\Sigma_\text{th}$ contains the exact dependence on the new physics Wilson coefficients. If the new physics Wilson coefficients are set to zero, it reduces to the theory covariance matrix in the SM. Expressing $\Sigma_\text{th}$ as above has the big advantage that the new physics dependence is given analytically and the time consuming numerical determination of $\Sigma_{\vec p}$ has to be performed only once.

In cases where the functions $f_k$ are non-trivial, the $\Sigma_\text{th}$ with the exact new physics dependence can not be found in a simple analytical way from $\Sigma_{\vec p}$. However, one can still find an analytic approximation in the limit of small new physics.
If the new physics Wilson coefficients are small compared to the SM values, we can expand the functions $f_k$ in $\epsilon$ and write them as polynomials
\begin{equation} \label{eq:approx}
 f_k(p_1,p_2,...,p_n) = p_k^\prime = \vec{p}_k^{\;\prime} \cdot \vec{V} + \mathcal O(\epsilon^3) = a_k^\prime +  \epsilon (\vec{b}_k^\prime \cdot \vec{C}) + \epsilon^2 (\vec{c}_k^{\;\prime} \cdot \vec{D}) + \mathcal O(\epsilon^3) ~.
\end{equation}
The coefficients of these polynomials are given by
\begin{equation}
 a_k^\prime = f_k(a_1,a_2,...,a_n) ~,\qquad \vec b_k^\prime = g^i_k \vec b_i ~, \qquad \vec c_k^{\;\prime} = g^i_k \vec c_i + \frac{1}{2} g^{ij}_k\, \text{vec}(\vec b_i \otimes \vec b_j) ~,
\end{equation}
where indices $i,j$ are summed over and we have defined the derivatives of $f_k(p_1, p_2, ..., p_n)$
\begin{equation}
 g_k^{i_1,i_2,...,i_\ell} =
 \frac{\partial^\ell\,f_k(p_1, p_2, ..., p_n)}{\partial p_{i_1}\partial p_{i_2}...\partial p_{i_\ell}}\Bigg|_{p_1=a_1, p_2=a_2, ..., p_n=a_n}\,.
\end{equation}
As above, it is straight forward to determine the covariance matrix $\Sigma_{\vec p^{\,\prime}}$ of the vectors $\vec p_k^{\;\prime}$. Since all approximated observables are linear in $\vec{p}_k^{\;\prime}$, we find analogously to~\eqref{eq:Sigma_th}
\begin{equation}  \label{eq:Sigma_th_approx}
 \Sigma_\text{th} = (\mathbb{1}_{N^\prime}\otimes \vec V^T) \cdot \Sigma_{\vec p^{\,\prime}} \cdot (\vec V \otimes \mathbb{1}_{N^\prime}) \Big|_{\epsilon = 1} ~,
\end{equation}
where $N^\prime$ is the number of polynomials $p_k^{\prime}$, which equals the number of observables.
The approximation can be improved systematically by expanding the functions $f_k$ in \eqref{eq:approx} to higher order in $\epsilon$. In that case, the vector $\vec V$ has to be extended to include higher powers of the Wilson coefficients. As the observables are still linear in the coefficients $\vec p_k^{\;\prime}$, \eqref{eq:Sigma_th_approx} continues to hold at any fixed order of the expansion. Note, however, that the size of the covariance matrix $\Sigma_{\vec p^{\,\prime}}$ grows rapidly with the order of the expansion.

\bibliographystyle{JHEP}
\bibliography{bibliography}

\end{document}